\input harvmac
\newcount\figno
\figno=0
\def\fig#1#2#3{
\par\begingroup\parindent=0pt\leftskip=1cm\rightskip=1cm\parindent=0pt
\global\advance\figno by 1
\midinsert
\epsfxsize=#3
\centerline{\epsfbox{#2}}
\vskip 12pt
{\bf Fig. \the\figno:} #1\par
\endinsert\endgroup\par
}
\def\figlabel#1{\xdef#1{\the\figno}}
\def\encadremath#1{\vbox{\hrule\hbox{\vrule\kern8pt\vbox{\kern8pt
\hbox{$\displaystyle #1$}\kern8pt}
\kern8pt\vrule}\hrule}}

\overfullrule=0pt

%
\def\underarrow#1{\vbox{\ialign{##\crcr$\hfil\displaystyle
 {#1}\hfil$\crcr\noalign{\kern1pt\nointerlineskip}$\longrightarrow$\crcr}}}
\def\tilde{\widetilde}

\def\inbar{\vrule height1.5ex width.4pt depth0pt}
\def\IC{\relax\hbox{\kern.25em$\inbar\kern-.3em{\rm C}$}}
\def\IR{\relax\hbox{\kern.25em$\inbar\kern-.3em{\rm R}$}}
\def\IZ{\relax\ifmmode\hbox{Z\kern-.4em Z}\else{Z\kern-.4em Z}\fi}

\font\zfont = cmss10 

\def\bigone{\hbox{1\kern -.23em {\rm l}}}
\def\ZZ{\hbox{\zfont Z\kern-.4emZ}}


\def\drawbox#1#2{\hrule height#2pt
        \hbox{\vrule width#2pt height#1pt \kern#1pt
              \vrule width#2pt}
              \hrule height#2pt}

\def\Asym#1#2{\vcenter{\vbox{\drawbox{#1}{#2}
              \kern-#2pt       
              \drawbox{#1}{#2}}}}

\batchmode
  \font\bbbfont=msbm10
\errorstopmode
\newif\ifamsf\amsftrue
\ifx\bbbfont\nullfont
  \amsffalse
\fi
\ifamsf
\def\IR{\hbox{\bbbfont R}}
\def\IC{\hbox{\bbbfont C}}

\def\IZ{\hbox{\bbbfont Z}}


\midinsert
\endinsert


\nref\weyl{H. Weyl, {\it Group Theory and Quantum Mechanics},
(Dover, New York, 1931)}

\nref\wigner{E.P. Wigner, Phys. Rev. {\bf 40}, 749 (1932).}

\nref\groenewold{A. Groenewold, Physica {\bf 12} (1946) 405-460.}

\nref\moyal{J.E. Moyal, Proc. Camb. Phil. Soc. {\bf 45}, 99
(1949).}

\nref\bayenone{F. Bayen, M. Flato, C. Fronsdal, A. Lichnerowicz
and D. Sternheimer, Ann. Phys. {\bf 111}, 61 (1978).}

\nref\bayentwo{F. Bayen, M. Flato, C. Fronsdal, A. Lichnerowicz
and D. Sternheimer,Ann. Phys. {\bf 111}, 111 (1978).}

\nref\reviews{ C.~K.~Zachos, ``A Survey of Star Product
Geometry,'' arXiv:hep-th/0008010; C.K. Zachos, ``Deformation
Quantization: Quantum Mechanics Lives and Works in Phase Space'',
Int. J. Mod. Phys. A {\bf 17} (2002) 297, hep-th/0110114; A.C.
Hirshfeld and P. Henselder, ``Deformation Quantization in the
Teaching for Quantum Mechanics'', Am. J. Phys. {\bf 70} (2002)
537; G. Dito and D. Sternheimer, ``Deformation Quantization:
Genesis, Developments and Metamorphoses, {\it Deformation
Quantization} (Strasbourg 2001) Lect. Math. Theor. Phys. {\bf 1}
Ed. de Gruyter, Berlin, IRMA (2002) pp. 9-54; D. Sternheimer,
``Deformation is Quantization'',
http://www.u-bourgogne.fr/monge/d.sternh/papers/DSoberwolfach.pdf.}

\nref\existen{M. De Wilde and P.B.A. Lecomte, Lett. Math. Phys.
{\bf 7} (1983) 478; H. Omori, Y. Maeda and A. Yoshioka, Adv. Math.
{\bf 85} (1991) 224.}

\nref\fedo{B. Fedosov, J. Diff. Geom. {\bf 40} (1994) 213.}

\nref\kont{M. Kontsevich, ``Deformation Quantization of Poisson
Manifolds I'', Lett. Math. Phys. {\bf 66} (2003) 157, e-Print:
q-alg/9709040.}

\nref\wwmformalism{R.L. Stratonovich, {\it Sov. Phys. JETP} {\bf
31}, 1012 (1956); G.S. Agarwal and E. Wolf, Phys. Rev. D {\bf 2}
(1970) 2161; 2206; A. Grossmann, {\it Commun. Math. Phys.} {\bf
48}, 191 (1976); J.M. Gracia-Bond\'{\i}a, Phys. Rev. A {\bf 30}
(1984) 691; J.M. Gracia Bond\'{\i}a and J.C. Varilly, {\it J.
Phys. A: Math. Gen.} {\bf 21}, L879 (1988), {\it Ann. Phys.} {\bf
190}, 107 (1989); J.F. Cari\~nena, J.M. Gracia Bond\'{\i}a and
J.C. Varilly, {\it J. Phys. A: Math. Gen.} {\bf 23}, 901 (1990):
M. Gadella, M.A. Mart\'{\i}n, L.M. Nieto and M.A. del Olmo, {\it
J. Math. Phys.} {\bf 32}, 1182 (1991);  M. Gadella, Fortschr.
Phys. {\bf 43} (1995) 229; J.F. Pleba\'nski, M. Przanowski and J.
Tosiek, {\it Acta Phys. Pol.} {\bf B27} 1961 (1996).}

\nref\hillery{M. Hillery, R.F. O'Connell, M.O. Scully and E.P.
Wigner, {\it Phys. Rep.} {\bf 106}, 121 (1984).}

\nref\tata{W.I. Tatarskii, Usp. Fiz. Nauk {\bf 139} (1983) 587.}

\nref\paco{
  G.~Dito and F.~J.~Turrubiates,
  ``The damped harmonic oscillator in deformation quantization,''
  Phys.\ Lett.\ A {\bf 352}, 309 (2006)
  [arXiv:quant-ph/0510150].}

\nref\hirshfeld{A.C. Hirshfeld and P. Henselder, ``Deformation
Quantization for Systems with Fermions'', Ann. Phys. (N.Y.) {\bf
302} (2002) 59.}

\nref\imelda{
  I.~Galaviz, H.~Garcia-Compean, M.~Przanowski and F.~J.~Turrubiates,
  ``Weyl-Wigner-Moyal Formalism for Fermi Classical Systems,'' to
  appear in Ann. Phys. (N.Y.), arXiv:hep-th/0612245.}

\nref\sw{N. Seiberg and E. Witten, ``String Theory and
Noncommutative Geometry'', JHEP {\bf 9909:03} (1999),
hep-th/9908142.}

\nref\dito{G. Dito, Lett. Math. Phys. {\bf 20} (1990) 125; Lett.
Math. Phys. {\bf 27} (1993) 73.}

\nref\curtright{T. Curtright D. Fairlie and C. Zachos, Phys. Rev.
D {\bf 58},  025002 (1998).}

\nref\zachos{ T. Curtright and C. Zachos, J. Phys. A {\bf 32}, 771
(1999).}

\nref\campos{H. Garc\'{\i}a-Compe\'an, J.F. Pleba\'nski, M.
Przanowski and F.J. Turrubiates, ``Deformation Quantization of
Classical Fields'', Int. J. Mod. Phys. A {\bf 16} (2001) 2533.}

\nref\strings{H. Garc\'{\i}a-Compe\'an, J.F. Pleba\'nski, M.
Przanowski and F.J. Turrubiates, ``Deformation Quantization of
Bosonic Strings'', J. Phys. A: Math. Gen. {\bf 33} (2000) 7935.}

\nref\imeldadirac{
  I.~Galaviz, H.~Garcia-Compean, M.~Przanowski and F.~J.~Turrubiates,
  ``Deformation Quantization of Fermi Fields,''
  arXiv:hep-th/0703125.}

\nref\antonu{ F. Antonsen, Phys. Rev. D {\bf 56}, 920 (1997);
``Deformation Quantization of Constrained Systems'',
gr-qc/9710021; ``Deformation Quantization of Gravity'',
gr-qc/9712012.}

\nref\LouisMartinezKW{
  D.~J.~Louis-Martinez,
  ``Weyl-Wigner-Moyal Formulation of a Dirac Quantized Constrained System,''
  Phys.\ Lett.\  A {\bf 269}, 277 (2000).}

\nref\KrivoruchenkoTGuno{
  M.~I.~Krivoruchenko, A.~A.~Raduta and A.~Faessler,
  ``Quantum Deformation of the Dirac Bracket'' Phys.\ Rev.\  D {\bf 73}, 025008
  (2006) [arXiv:hep-th/0507049].}

\nref\KrivoruchenkoTGdos{
  M.~I.~Krivoruchenko,
  ``Moyal Dynamics of Constraint Systems'', [arXiv:hep-th/0610074].}

\nref\deriglazov{
  A.~A.~Deriglazov, ``Noncommutative Relativistic Particle on the Electromagnetic
  Background,'' Phys.\ Lett.\  B {\bf 555}, 83 (2003), [arXiv:hep-th/0208201].}

\nref\hori{T.~Hori, T.~Koikawa and T.~Maki, ``Moyal Quantization
for Constrained System,'' Prog.\ Theor.\ Phys.\  {\bf 108}, 1123
(2002) [arXiv:hep-th/0206190].}

\nref\IsidroXR{
  J.~M.~Isidro, ``Dirac brackets from magnetic backgrounds,''
  arXiv:hep-th/0611026.}

\nref\dirac{ P.A.M. Dirac, Can. J. Math. {\bf 2} (1950) 129; {\it
Lectures on Quantum Mechanics}, (Belfer Graduate School of
Science, Yeshiva University, New York, 1964.)}

\nref\firstconstr{A. Hanson, T. Regge and C. Teitelboim, {\it
Constrained Hamiltonian Systems}, (Accademia Nazionale dei Lincei,
Rome, 1976).}

\nref\sunder{K. Sundermeyer, {\it Constrained Dynamics}, Springer
Lecture Notes in Physics Vol. 169 (Springer-Verlag,
Berlin-Heidelberg-New York, 1982).}

\nref\hennteit{M. Henneaux and C. Teitelboim, {\it Quamtization of
Gauge Systems}, Princeton University Press, Princeton NJ, 1990.}

\nref\gt{D.M. Gitman and I.V. Tyutin, {\it Quantizations of Fields
and Constraints}, Springer-Verlag, Heidelberg (1990).}

\nref\gitman{D.M. Gitman and I.V. Tyutin, {it Quantizations of
Fields and Constraints}, Springer-Verlag, Heidelberg (1990) pp.
229-253; Class. Quantum Grav. {\bf 7} (1990) 2131.}

\nref\ps{P.S. Gavrilov and D.M. Gitman, Class. Quantum Grav. {\bf
10} (1993) 57.}

\nref\tuckey{J.M. Evans and P.A. Tuckey, ```A Geometrical Approach
to Time-dependent Gauge-fixing'', Int. J. Mod. Phys. A {\bf 8}
(1993) 4055, arXiv:hep-th/9208009.}

\nref\evans{J.M. Evans, ```Time Independent Gauge Fixing and the
Quantization of a Particle in a General Electromagnetic
Background'', Class. Quantum Grav. {\bf 10} (1993) L221.}

\nref\gsw{ M.B. Green, J.H. Schwarz and E. Witten, {\it Superstring Theory},
Two volumes, Cambridge Univesity Press, Cambridge (1986).}

\nref\pol{ J. Polchinski, {\it String Theory}, Two volumes, Cambridge University
Press, Cambridge (1998).}

\nref\theisen{ D. L\"ust and S. Theisen, {\it Lectures on  String
Theory}, Lecture Notes in Physics {\bf 346}, Springer-Verlag,
Berlin (1989).}

\nref\barton{B. Zwiebach, {\it A First Course in String Theory},
Cambridge University Press (2005).}

\nref\naka{T. Maskawa and H. Nakajima, ``Singular Lagrangian and
the Dirac-Faddeev Method'', Prog. Theor. Phys. {\bf 56} (1976)
1295.}

\vskip -3.5 truecm
\Title{CINVESTAV-FIS-07/15}
{\vbox{\centerline{Deformation Quantization of Relativistic
Particles}
\medskip
\centerline{in Electromagnetic Fields}}\foot{This work represents
a thesis for a Bachelor degree in Physics submitted at the
Facultad de Ciencias, UAE, Toluca, M\'exico by Laura S\'anchez. We
dedicate this paper to the Memory of Prof. Guillermo Moreno.}}

\vskip -.5truecm

\centerline{Laura S\'anchez$^{a}$, Imelda
Galaviz$^{b}$\foot{E-mail address: {\tt
igalaviz@fis.cinvestav.mx}} and Hugo
Garc\'{\i}a-Compe\'an$^{b,c}$\foot{E-mail address: {\tt
compean@fis.cinvestav.mx}}}
\smallskip

\centerline{\it $^a$Facultad de Ciencias, U.A.E. M\'exico}
\centerline{\it Instituto Literario 100, 50000, Toluca, M\'exico}

\centerline{\it $^b$Departamento de F\'{\i}sica} \centerline{\it
Centro de Investigaci\'on y de Estudios Avanzados del IPN}
\centerline{\it Apdo. Postal 14-740, 07000, M\'exico D.F.,
M\'exico}

\centerline{\it $^c$Centro de Investigaci\'on y de Estudios
Avanzados del IPN, Unidad Monterrey} \centerline{\it Cerro de las
Mitras 2565, cp. 64060, Col. Obispado, Monterrey N.L., M\'exico}


\baselineskip 18pt
\medskip
\noindent The Weyl-Wigner-Moyal formalism for Dirac second class
constrained systems has been proposed recently as the deformation
quantization of Dirac bracket. In this paper, after a brief review
of this formalism, it is applied to the case of the relativistic
free particle. Within this context, the Stratonovich-Weyl
quantizer, Weyl correspondence, Moyal $\star$-product and Wigner
function in the constrained phase space are obtained. The recent
Hamiltonian treatment for constrained systems, whose constraints
depend explicitly on time, are used to perform the deformation
quantization of the relativistic free charged particle in an
arbitrary electromagnetic background. Finally, the system
consisting of a charged particle interacting with a dynamical
Maxwell field is quantized in this context.

\noindent PACS numbers: {\it 11.10.-z, 11.10.Ef, 03.65.Ca,
03.65.Db}

\noindent
Key words: {\it Deformation Quantization, Relativistic
Particles, Electromagnetic Fields}

\Date{May, 2007}

\newsec{Introduction}

At the present time deformation quantization constitutes an
alternative method to quantize classical systems. It started with
the Weyl correspondence in quantum mechanics between classical
observables as the algebra of smooth functions ${\cal A}_\Gamma$
on the phase space $\Gamma$ and linear operators acting over
certain Hilbert space ${\cal H}$
\refs{\weyl,\wigner,\groenewold,\moyal}. This is a one-one
correspondence where the (noncommutative) operator product is
mapped to an associative and non-commutative product, called the
Moyal star product $\star$. This product is defined on the formal
series-valued ring of functions ${\cal A}_\Gamma[[\hbar]]$ on
$\Gamma$. This correspondence is today known as the
Weyl-Wigner-Moyal (WWM) correspondence or WWM formalism.

The theory of deformation quantization was formulated in more
rigorous mathematical setting in the context of deformation theory
in Refs. \refs{\bayenone,\bayentwo}. In particular, in Ref.
\bayentwo\ there were described some examples of quantum systems
in terms of deformations of the symplectic structure. For
instance, the spectrum of the hydrogen atom was computed
explicitly. Thus the deformation quantization proved to be
equivalent to another quantization methods as the canonical
quantization or the Feynman path integrals. This equivalence seems
to point out the fact that {\it deformation is quantization}.
Deformation quantization from the physical point of view has been
be surveyed recently in Refs. \refs{\reviews}.

Later there was found that star product in fact does exist for any
symplectic manifold \refs{\existen}. Moreover, there was an
explicit construction, trough the uses of symplectic differential
geometry, due Fedosov \fedo. More recently, Kontsevich proved that
the star product, in general, does exist for any Poisson manifold
\kont.

The most part of the work on deformation quantization has been
carried over for the years to dynamical systems with a finite
number of degrees of freedom \refs{\wwmformalism,\hillery,\tata}.
One of the recent applications is, for instance, to the damped
oscillator \paco. The other is the extension of deformation
quantization to classical systems with a finite number of
fermionic degrees of freedom (Grassmann variables)
\refs{\hirshfeld,\imelda}. More recently the Weyl correspondence
has been playing an important role in the context of open string
theory in the presence of a $B$-field \sw.

However, the method of deformation quantization also has been
applied to describe quantum fields and strings. To be more
precise, in Ref. \dito, the problem of the UV divergences in the
vacuum energy was discussed. Further developments can be found in
Refs. \refs{\curtright,\zachos,\campos}. Furthermore, the
quantization of bosonic strings was discussed later in Ref.
\strings. The extension of deformation quantization to infinite
number of fermionic degrees of freedom was studied recently in
Ref. \imeldadirac.

Also it has been applied to gravitational systems and systems with
constraints \refs{\antonu}. In particular in the first reference,
dynamical systems with first class constraints were studied. In
the present paper we continue with this philosophy and we will
quantize the relativistic particle for the free and interacting
cases. In order to do that we will use the recent results
\refs{\LouisMartinezKW,\KrivoruchenkoTGuno,\KrivoruchenkoTGdos},
concerning the WWM formalism for a second class constrained
system. For related results on this subject, see Refs.
\refs{\deriglazov,\hori,\IsidroXR}. Within this prescription of
quantization for second class constraints, once we gauge fixing,
we can re-express all the formalism in the reduced phase space in
terms of the physical variables only. We shall observe that the
description can be obtained from the low energy limit $\alpha
'=\ell_S^2 \to 0,$ of the deformation quantization of the bosonic
string worked out at \strings.

In general terms, the description of classical constrained systems
(for classical reviews, see
\refs{\dirac,\firstconstr,\sunder,\hennteit,\gt}) with constraints
depending explicitly on time represent a challenge, even at the
classical level. A proposal which changes radically the
Hamiltonian form of the theory was introduced in Refs.
\refs{\gitman,\ps}. Recently, a geometrical proposal which
preserved the Hamiltonian evolution equation presented on the
constrained phase space, also to the physical (reduced) phase
space, was done in Ref. \tuckey. This treatment allowed to
quantize the second class constrained system of the relativistic
charged particle moving in an arbitrary electromagnetic background
\evans. Precisely this procedure is what allows to find a
Hamiltonian formulation on the physical phase space in order to
prove the consistency of the deformation quantization for second
class constrained systems proposed in
\refs{\LouisMartinezKW,\KrivoruchenkoTGuno,\KrivoruchenkoTGdos}.
In particular, we use this formalism to quantize by deforming the
phase space of a free relativistic particle, in an arbitrary
electromagnetic background and interacting with a dynamical
electromagnetic field.

Our paper is organized as follows. In Sec. 2 we overview the
covariant description of a relativistic free particle. Sec. 3 is
devoted to give in general the WWM formalism on the constrained
phase space. Thus the Stratonovich-Weyl (SW) quantizer, which is
the main object to determine the Weyl correspondence, is found.
The Moyal $\star$-product, Wigner function, some of its properties
and correlation functions are also defined. In Sec. 4, we show
that the formalism to deal time-dependent second class constraints
given at \refs{\tuckey,\evans} are precisely what is needed to
perform appropriately the deformation quantization in the
constrained phase space from Refs.
\refs{\LouisMartinezKW,\KrivoruchenkoTGuno,\KrivoruchenkoTGdos}.
All ingredients of the WWM are computed in the light-cone gauge
and it is shown that they coincides with the low-energy effective
theory with $\alpha ' \to 0$ from the bosonic string \strings.
Some properties of the Wigner function concerning its
correspondence with a pure state is discussed in an appendix. Also
in Sec. 5 the deformation quantization of the particle in an
arbitrary electromagnetic background is also described within this
formalism. In Sec. 6, we discuss the deformation quantization of
the relativistic charged particle interacting with a dynamical
electromagnetic field. This is done in the Lorentz and ligh-cone
gauge. Finally, in Sec. 7, some concluding remarks close the
paper.

\vskip 2truecm
\newsec{Brief Overview on Relativistic Particles}

In this section we give an overview of the theory of relativistic
particles. Our aim is not to provide an extensive review of such
theory, but briefly to recall the notation and conventions, which
will be strictly necessary in the following sections (for further
details see \refs{\gsw,\pol,\theisen,\barton}). In particular we
will follow notation and conventions from Ref. \barton. In
particular, we will take in our analysis $c=1$.

To perform the description we consider the world-line $L$ embedded
into the $D$-dimensional space-time $M$ of Lorentzian metric
$\eta_{\mu \nu} = diag(-1,1,\dots ,1),$ $\mu, \nu=0,1, \dots
,D-1$. This embedding is defined by the set of functions: $X^{\mu}
= X^{\mu} (\tau),$ where $\tau$ is the coordinate on the
world-line $L$. The dynamics is encoded in the Nambu-Goto action $
S_{NG} =\int_L d \tau L_{NG}= -m \int_L d \tau \sqrt{- \eta_{\mu
\nu} {dX^\mu \over d \tau} {dX^\nu \over d \tau}},$ where $m$ is
the mass of the relativistic particle.

Let $e(\tau)$ be a world-line metric on $L$. The dynamics of the
scalar fields $X^{\mu}$ is described by a classically equivalent
action to the Nambu-Goto action by:
$$
S_P = \int_L d \tau L_{P}
$$
\eqn\dos{ = \int_{L} d \tau \bigg[{e^{-1}(\tau)\over 2} \eta_{\mu
\nu}{dX^{\mu} \over d \tau} {dX^{\nu}\over d\tau} - {e \over
2}m^2\bigg]. }  While the constraint $T={\delta S_P \over \delta
e}=0$ is given by \eqn\ocho{ T =  {e^{-2}\over 2}\eta_{\mu \nu}
{dX^{\mu} \over d \tau} {dX^{\nu}\over d\tau} + {m^2\over 2} = 0.}

In the conformal gauge $e(\tau)={1\over m^2}$ then the action is:
\eqn\dospuntocinco{ S_P = \int_{L} d \tau \bigg[{m^2\over 2}
\eta_{\mu \nu}{dX^{\mu} \over d \tau} {dX^{\nu}\over d\tau} -
{1\over 2}\bigg]. } Then, the equations of motion of the free
particle are the usual ones \eqn\tres{{d^2X^{\mu} \over d \tau^2}
= 0.} The general solution of Eqs. \tres\ can be written in the
form

\eqn\doce{
 X^{\mu}(\tau) = x^{\mu} + {1 \over m^2}p^{\mu}\tau, }
where $x^{\mu}$ and $p^{\mu}$ are real variables. The canonical
momentum $\Pi^{\mu}$ of $X^{\mu}$ is as usual defined by $\Pi_\mu
= {\partial {L}_P \over \partial \dot{X}^\mu}$. This is given by

\eqn\trece{ \Pi^{\mu}(\tau) = m^2 \dot{X}^\mu =p^{\mu}.} These
solutions $(X^{\mu},\Pi^{\mu})$ satisfy the standard Poisson
brackets

$$\{ X^{\mu}(\tau), \Pi^{\nu}(\tau) \}_P
= \eta^{\mu \nu},
$$
\eqn\quince{
 \{ X^{\mu}(\tau), X^{\nu} (\tau) \}_P = 0 =
\{ \Pi^{\mu} (\tau), \Pi^{\nu}(\tau) \}_P.} Poisson brackets for
variables $x^{\mu}, \ p^{\mu}$ are

\eqn\dseis{ \{x^{\mu}, p^{\nu} \}_P = \eta^{\mu \nu},} with the
remaining independent Poisson brackets being zero.

The constraint \ocho\ is written in the conformal gauge as:
\eqn\primary{ T =  {1\over 2}\big[p^2 + m^2 \big] = 0.} This is a
primary first class constraint which generate the
$\tau$-reparametrization invariance.

Remember that $\Phi_1= p^2+m^2=0$ is the constraint imposed by the
on-shell condition. In the {\it light-cone gauge} the constraint
equations \ocho\ can be easily solved and then eliminated.  This
gauge will be crucial for the deformation quantization of the
relativistic particle in order to identify the relevant phase
space where implement this quantization.

First, introduce the light-cone coordinates $ X^{\pm} := {1 \over
\sqrt{2}}( X^0 \pm X^{D-1}),$ and the remaining transverse
coordinates $\vec{X}_T= X^j,$ and  $\vec{\Pi}_T= \Pi^j,$ where
$j=1, \dots , D-2$ are left as before. As $X^+(\tau)$ satisfies
the equation of motion (2.2) one can choose the coordinate $\tau$
in such a manner that $X^+(\tau) = {1 \over m^2} p^+ \tau. $ Of
course, this gauge fixing condition $X^+(\tau) ={1 \over m^2} p^+
\tau$ is explicitly time-dependent. Later we are going back to
describe a procedure to deal such a situation (see
\refs{\tuckey,\evans}).

In this gauge we can solve the constraint equations in the sense
that $p^-$, are determined totally in terms of $p^+$ and $p^j$.
Thus the number of degrees of freedom i.e., the independent
dynamical variables of the relativistic particle after imposing
constraints and gauge conditions are $(x^-,p^+,
\vec{X}_T,\vec{\Pi}_T)$, or, equivalently: $(x^-,p^+,\vec{x}_T,
\vec{p}_T)$. Thus we have $D-1$ degrees of freedom. For the
Poisson bracket for these variables we have
$$
\{x^-,p^+\}_P = -1, \ \ \ \  \ \ \ \ \ \
\{X^j(\tau),\Pi^k(\tau)\}_P: = \delta^{jk},
$$
\eqn\poissonone{ \{X^j(\tau),X^k(\tau)\}_P=0=
\{\Pi^j(\tau),\Pi^k(\tau)\}_P.}

In general, the total Hamiltonian $H_T$ can be written as the sum
of the canonical Hamiltonian $H_C$ plus primary first class
constraint $\Phi_1(X^\mu,\Pi_\mu)$ as follows \eqn\hamiltot{ H_T =
H_C +  \lambda_1 \Phi_1(X^\mu,\Pi_\mu),} where $\lambda_1$ is an
arbitrary function (Lagrange multipliers).  Remember that $H_C$ is
vanishing for every reparametrization invariant system. After
imposing the light-cone gauge the Hamiltonian is given by
\eqn\ham{ H_T={p^+ p^- \over m^2}= {1 \over 2m^2}
\sum_{j=1}^{D-2}(p_j^2 + m^2).} In the next section we study the
problem of deformation quantization more systematically from the
point of view of constrained systems.

\newsec{Deformation Quantization of Second Class Constrained Systems}

In the present section we will overview the WWM formalism
\refs{\wwmformalism,\hillery,\tata} for the case of general
systems with second class constraints
\refs{\LouisMartinezKW,\KrivoruchenkoTGuno,\KrivoruchenkoTGdos}.
The description includes a deformation of the Dirac bracket and
depends on the local considerations of the physical (or reduced)
phase space.

\vskip 1truecm
\subsec{Preliminaries of Constrained Systems}

It is well known that the dynamics of the relativistic particle
constitutes an example of a first class constrained system.
However it also can be regarded as a second class constrained
system, when the gauge conditions are taken into account
\refs{\hennteit,\gt}).

In order to implement the quantization one can adopt two possible
positions. The first one is simply to impose the constraints and
gauge fixing conditions at the classical level, and quantizing
only the relevant (physical) degrees of freedom. This has been
done previously for the deformation quantization of the bosonic
string in the light-cone gauge \refs{\strings}. The second
possibility is to develop the formalism of deformation
quantization specifically before imposing constraints and gauge
fixing. This formalism was developed recently in Refs.
\refs{\LouisMartinezKW,\KrivoruchenkoTGuno,\KrivoruchenkoTGdos}.
In the present section we will overview some material from these
constructions. In the next sections we will apply them to the
relativistic free particle and the charged particle inside an
electromagnetic background and its interaction with a dynamical
electromagnetic field.

The phase space before imposing constraints and gauge fixing is
given by the so called constrained phase space which we assume of
the form: ${\cal Z}_P = \{Z^\alpha=(X^\mu,\Pi_\mu) \in \IR^{2D}\}
= T^*\IR^D \cong \IR^{2D}$. Here $Z^\alpha$ are the coordinates of
the extended phase space: $Z^\alpha=X^\mu$ for $\alpha=1, \dots,
D$ and $Z^\alpha=\Pi_\mu$ for $\alpha=D+1, \dots, 2D$. The
corresponding symplectic two-form is given by $\omega =
\omega_{\alpha \beta} dZ^\alpha \wedge dZ^\beta$. Due Darboux's
theorem always is possible to find the symplectic basis where the
entries of $\omega_{\alpha \beta}$ are constant and moreover
$\omega_{\alpha \beta} = \pmatrix{0 & {\bf 1}_{D \times D} \cr
-{\bf 1}_{D \times D} & 0}.$ Therefore the symplectic form takes
the canonical form $\omega= dX^\mu \wedge d \Pi_\mu.$

From the second class constraints perspective, first class
constraints plus gauge fixing conditions can be regarded as a
second class constraint system \refs{\hennteit,\gt}. Thus we have
a set of purely second class constraints: $\Phi_I(Z^\alpha)$ such
that $\det(C_{IJ})=\det(\{\Phi_I,\Phi_J \}_P) \not= 0$ with $I=1,
\dots ,M$, where $M$ as to be even, i.e., $M=2m$. For the
relativistic particle these two second class constraints are
$\Phi_1=p^2+m^2$ and $\Phi_2= X^+-{1\over m^2}p^+\tau.$

Therefore a constrained submanifold $\Sigma_I$ can be associated
to each constraint $\Phi_I$. Thus the dynamics of the system is
described through the intersection of all constraint submanifolds
in the form ${\cal Z}_\cap= \Sigma_1 \cap \cdots \cap \Sigma_M$.
By a theorem of Maskawa and Nakajima \naka, we can identify ${\cal
Z}_\cap$ with the reduced phase space ${\cal Z}_P^{\cal R}$. Thus
in general the constrained submanifold is ${\cal Z}_\cap={\cal
Z}_P^{\cal R}= \IR^{2(D-m)}$. For the case of the relativistic
particle we have two second class constraints ($m=1$) and
therefore the reduced phase space is ${\cal Z}_P^{\cal R}=
\IR^{2(D-1)}$. Thus the number of physical degrees of freedom is
$D-1$, i.e., the transverse degrees of freedom $\vec{X}_T$ plus
$x^-$.

One of the features of these kind of systems is that the matrix
generated by the constraints: $\det{({\bf C})}= \det{(
C_{IJ})}\not=0$ is the main object in the description. Therefore
there exist the inverse matrix $C^{-1}_{IJ}:=C^{IJ}=-C_{IJ}$. The
standard procedure establishes that the Poisson bracket on the
constraint manifold has to be changed to the Dirac bracket in the
extended phase space \eqn\diracparen{ \{f,g\}_D = \{f,g\}_P -
\sum_{I,J=1}^{M}\{f,\Phi_I\}_P C^{-1}_{IJ} \{\Phi_J,g\}_P,} for
any pair of functions $f(Z^\alpha)$ and $g(Z^\alpha)$. With the
known property $\{f,g\}_D \buildrel{{\cal Z}_P^{\cal R}}\over{=}
\{f,g\}_P$. The Dirac bracket satisfies the same properties as the
Poisson bracket.

Similarly as the Poisson bracket, the Dirac bracket also can be
expressed in geometrical terms \eqn\invforma{ \omega^{-1}_D(df,dg)
= \{f,g\}_D =\omega^{\alpha \beta}_D {\partial f \over \partial
Z^\alpha} \cdot {\partial g \over \partial Z^\beta},} where we
have \eqn\formainverse{ \omega^{\alpha \beta}_D=
\{Z^\alpha,Z^\beta\}_{D}.}

\vskip 1truecm \subsec{Skew-gradient Projection Method}

Now locally, in the symplectic basis, one has
$\{\Phi_I(Z^\alpha),\Phi_J(Z^\alpha)\}_D = I_{IJ}$ with $I,J=1,
\dots , 2m$, where $I_{IJ} = \pmatrix{0 & {\bf 1}_{m \times m} \cr
-{\bf 1}_{m \times m} & 0}.$ Then the restriction over the reduced
phase space ${\cal Z}_P^{\cal R}$ starting from the constrained
phase space ${\cal Z}_P$ can be represented through a
skew-gradient projection $Z_S$ of the canonical coordinates
$Z^\alpha$ onto ${\cal Z}_P^{\cal R}$
\refs{\KrivoruchenkoTGuno,\KrivoruchenkoTGdos}

\eqn\project{ Z_S(Z^\alpha)= Z^\alpha + X^I \Phi_I(Z^\alpha) + {1
\over 2!} X^{IJ} \Phi_I(Z^\alpha) \Phi_J(Z^\alpha) + \cdots . }
Such projection is done according to the condition of having not
variations along the phase space flows generated by the
constraints $\Phi_I$, i.e., \eqn\flujo{
\{Z_S(Z^\alpha),\Phi_I(Z^\alpha)\}_P=0,} which implies that
$\Phi_I(Z_S(Z^\alpha))=0$. i.e. $Z_S(Z^\alpha)$ lies on the
reduced phase space.

 This construction has the nice property that any function $f(Z^\alpha)$ on
the phase space ${\cal Z}_P$ can be projected on ${\cal Z}_P^{\cal
R}$ as follows \eqn\defini{ f_S(Z^\alpha)= f(Z_S(Z^\alpha)),} and
satisfies \eqn\flujoproj{ \{f_S(Z^\alpha),\Phi_I(Z^\alpha)\}_P=0.}
On the reduced phase space we have $Z^\alpha_S(Z^\alpha)=
Z^\alpha$ and any observable $f$ can be replaced by $f_S$. These
projected functions satisfy the properties:
$\{f,g\}_D=\{f_S,g\}_P=\{f,g_S\}_P= \{f_S,g_S\}_P$. This
projection also defines an equivalence class of functions as
follows: $[f]=\{f(Z^\alpha)\sim g(Z^\alpha), \ \ {\rm iff} \ \
f_S(Z^\alpha)\buildrel{{\cal Z}_P^{\cal R}}\over{=}
g_S(Z^\alpha)\}$. Thus the manner of incorporating this in the WWM
formalism is that the integrations $\int_{{\cal Z}_P} F(Z^\alpha)
d^{2D}Z$ over the constrained phase space ${\cal Z}_P$ should be
localized over the space of equivalence classes or simply over the
reduced phase space. Therefore, all integrations over the extended
phase space should be localized at ${\cal Z}_P^{\cal R}$

\eqn\localizacion{ \int_{{\cal Z}_P} \delta({\cal Z}_P^{\cal R})
F(Z^\alpha) d^{2D}Z= \int_{{\cal Z}_P^{\cal R}}
F(Z_S)d^{2(D-m)}Z.} Later we will go back to discuss this subject.

The dynamics of the systems are controlled by the Dirac bracket
and the evolution equation is given by

\eqn\dinamica{ \{f,h\}_D = {\partial f \over \partial t},} for a
given Hamiltonian function $h$ and for any function $f$. The
projection to the reduced phase space reads \eqn\dinamicados{
\{f,h_S\}_P = {\partial f \over \partial t},} with $h_S=h(Z_S).$

\vskip 1truecm \subsec{WWM Formalism in the Constrained Phase
Space}

The formulation in the extended phase space needs from the action
of operators on the extended Hilbert space ${\cal H}_e$
$$
\widehat{X}^\mu |X^\mu \rangle = X^\mu |X^\mu \rangle , \ \ \ \ \
\ \widehat{\Pi}^\mu |\Pi^\mu \rangle = \Pi^\mu |\Pi^\mu \rangle ,
$$
\eqn\scuatro{ [\widehat{X}^\mu,\widehat{\Pi}^\nu] = i \hbar
\eta^{\mu \nu}.} With these definitions we set the normalization
of these states as follows

\eqn\sseis{ \int d^{D} X | X^\mu\rangle \langle X^\mu| =
\widehat{1} \ \ \ \ \ {\rm and} \ \ \ \ \ \int  d^{D}({\Pi \over 2
\pi \hbar}\big) |\Pi^\mu\rangle \langle \Pi^\mu| = \widehat{1}.}

Now we define the {\it Stratonovich-Weyl ({\rm SW}) quantizer} as
follows
$$
\widehat{\Omega}(Z^\alpha) = \int d^{D} \xi \exp \bigg\{ -{i\over
\hbar} \xi^\mu \Pi_\mu \bigg\} \bigg|X^\mu - {\xi^\mu \over 2}
\bigg{\rangle} \bigg{\langle} X^\mu + {\xi^\mu \over 2}\bigg|
$$
\eqn\socho{ = \int d^{D}({\eta \over 2 \pi \hbar}) \exp \bigg\{
-{i \over \hbar} \eta^\mu X_\mu \bigg\} \bigg|\Pi^\mu + {\eta^\mu
\over 2} \bigg\rangle \bigg\langle \Pi^\mu - {\eta^\mu \over
2}\bigg|. } The SW quantizer has the important properties:
$$
\big( \widehat{\Omega}(Z^\alpha)\big)^{\dag} =
\widehat{\Omega}(Z^\alpha), \ \ \ \ \ \ \ \ \ {\rm Tr}
\big(\widehat{\Omega}(Z^\alpha)\big) = 1,
$$
\eqn\snueve{ {\rm Tr} \big[\widehat{\Omega}(Z^\alpha)
\widehat{\Omega}(Z'^\alpha)\big] = \delta(Z^\alpha - Z'^\alpha).}
Here $\delta(Z^\alpha - Z'^\alpha)= \delta(X^\mu - X'^\mu) \delta
\big({\Pi^\mu - \Pi'^\mu \over 2 \pi \hbar} \big).$ Here ${\rm
Tr}$ is the trace over the extended Hilbert space ${\cal H}_e$.

In the Hilbert space representation the quantization condition
reads $\{\cdot,\cdot\}_D \to {1 \over i \hbar}[\cdot,\cdot].$ Then
we have $[\widehat{\Phi}_I(Z^\alpha),\widehat{\Phi}_J(Z^\alpha)] =
i \hbar \widehat{I}_{IJ}$ with $I=1, \dots , 2m$. Then the quantum
version of the restriction over the reduced phase space ${\cal
Z}_P^{\cal R}$ is given by the quantum version of the
skew-gradient projection $\widehat{Z}_S$

\eqn\projectone{ \widehat{Z}_S(\widehat{Z}^\alpha)=
\widehat{Z}^\alpha + X^I \widehat{\Phi}_I(\widehat{Z}^\alpha) + {1
\over 2} X^{IJ} \widehat{\Phi}_I(\widehat{Z}^\alpha)
\widehat{\Phi}_J(\widehat{Z}^\alpha) + \cdots . } Such a
projection is also done respecting the condition of having not
variations along the phase flows generated by the operator
constraints $\widehat{\Phi}_I$, i.e., \eqn\flujoquantum{
[\widehat{Z}_S,\widehat{\Phi}_I(\widehat{Z}^\alpha)]=0,} which
implies that $\widehat{\Phi}_I(\widehat{Z}_S(Z^\alpha))=0$. i.e.
$\widehat{Z}_S(\widehat{Z}^\alpha)$ lies on the quantum reduced
phase space.

In this construction one has that any operator-valued function
$\widehat{f}(\widehat{Z}^\alpha)$ on the phase space is projected
on the reduced phase space as follows

\eqn\definioperat{ \widehat{f}_S(\widehat{Z}^\alpha)=
\widehat{f}(\widehat{Z}_S({Z}^\alpha)),} and satisfies
\eqn\flujoprojops{
[\widehat{f}(\widehat{Z}_S(Z^\alpha)),\widehat{\Phi}_I(Z^\alpha)]=0.}
On the reduced phase space we have
$\widehat{Z}^\alpha_S(Z^\alpha)= \widehat{Z}^\alpha$ and therefore
any observable $\widehat{f}$ can be replaced $\widehat{f}_S$.
These projected operators satisfy the following properties:
$[\widehat{f},\widehat{g}]=[\widehat{f}_S,\widehat{g}]=[\widehat{f},\widehat{g}_S]=
[\widehat{f}_S,\widehat{g}_S]$. This projection also defines an
equivalence class of operators \eqn\equivalenceclass{
[\widehat{f}]=\{\widehat{f}(Z^\alpha)\sim \widehat{g}(Z^\alpha), \
\ {\rm iff} \ \ \widehat{f}_S(Z^\alpha)\buildrel{{\cal Z}_P^{\cal
R}}\over{=} \widehat{g}_S(Z^\alpha)\}.}

The quantum dynamics is described by the evolution equation
\eqn\qdinamica{[\widehat{f},\widehat{h}_S] = i \hbar {d \over
dt}\widehat{f},} where $\widehat{h}_S$ is the skew-gradient
projection of $\widehat{h}$.

As we will see, Weyl correspondence is then carried over
equivalence classes such that it identifies an equivalence class
of functions with the corresponding equivalence class of
operators. This is a one-one correspondence.

Let $F = F(Z^\alpha)$ be a function on the extended phase space
${\cal Z}_P$\foot{Actually $F$ is a formal series in $\hbar$ such
that $F(Z^\alpha,\hbar)$ is an element of $C^\infty({\cal
Z}_P)[[\hbar]]$, the ring of formal series in $\hbar$ with values
in the smooth functions over ${\cal Z}_P$.}. Then according to the
Weyl rule \refs{\wwmformalism} one assign the following operator
$\widehat{F}$ corresponding to the equivalence class of symbols:
$[F] = \{ F(Z^\alpha)\sim F'(Z^\alpha), \ F_S(Z^\alpha)
\buildrel{{\cal Z}_P^{\cal R}}\over {=} F'_S(Z^\alpha) \}.$ Thus
we have

\eqn\weylc{ \widehat{F} = W(F)= \int \ d^{2D}Z \sqrt{\det {\bf C}}
\prod_{I=1}^{2m} \delta[\Phi_I(Z^\alpha)]  \ F(Z^\alpha)
\widehat{\Omega}(Z^\alpha),} where $d^{2D}Z= d^{D} X d^{D}( {\Pi
\over 2 \pi \hbar}).$ $F$ is called the $symbol$ of the operator
$\widehat{F}$, $Sym(\widehat{F})=F(Z^\alpha)$.

After the constraints are imposed through the integration over the
$\delta$'s in Eq. \weylc\ we have \eqn\weylrule{ \widehat{F} =
W(F)= \int \ d^{2(D-m)}Z \sqrt{\det {\bf C}} \ F_S(Z^\alpha)
\widehat{\Omega}_S(Z^\alpha),} where
$\widehat{\Omega}_S(Z^\alpha)$ is the Stratonovich-Weyl quantizer
restricted to ${\cal Z}_P^{\cal R}$. Also we have
$\widehat{\Omega}(Z^\alpha)$ and taking the trace one has

\eqn\sedos{ F(Z^\alpha) = W^{-1}(\widehat{F}) = {\rm Tr} \bigg(
\widehat{\Omega}(Z^\alpha) \widehat{F} \bigg).}

Now, let $F = F(Z^\alpha)$ and  $G = G(Z^\alpha)$ be elements of
$C^\infty({\cal Z}_P)[[\hbar]]$ and let $\widehat{F}= W(F)$ and
$\widehat{G} =W(G)$ be their corresponding operators. We would
like to find what function on ${\cal Z}_P$ corresponds to the
product $\widehat{F} \widehat{G}.$ This function is denoted by $F
\star_D G$ and it is called the {\it Moyal $*$-product of} $F$ and
$G.$

By using the properties from Eq. \sedos\  one gets

\eqn\setres{ \big(F \star_D G \big) (Z^\alpha) =
W^{-1}(\widehat{F} \widehat{G})= {\rm Tr} \big[
\widehat{\Omega}(Z^\alpha) \widehat{F} \widehat{G} \big].}
Substituting Eq. \weylc\ into \setres, using then \socho\ and
performing straightforward but tedious manipulations we finally
obtain \eqn\proddemoyal{ \big(F \star_D  G \big) (Z^\alpha)=
F(Z^\alpha_1) \exp\bigg\{{i\hbar\over 2}
\buildrel{\leftrightarrow}\over {\cal P}_D\bigg\} G(Z^\alpha_2)
\bigg|_{Z_1^\alpha=Z_2^\alpha=Z^\alpha},} where \eqn\diracoper{
\buildrel{\leftrightarrow}\over {\cal P}_D = \omega^{\alpha
\beta}_D {\buildrel{\leftarrow}\over{\partial} \over \partial
Z^\alpha} \cdot {\buildrel{\rightarrow}\over{\partial} \over
\partial Z^\beta}.}

The Moyal bracket associated to the Dirac bracket is given by
\eqn\proddedirac{ \{F,G\}^{\star_D}_M= {1 \over i \hbar}(F \star_D
G - G \star_D F).}

Thus, it constitutes a quantum deformation of the Dirac bracket.
This can be regarded also as an extension of the Moyal bracket to
second class constrained systems. Thus we have \eqn\deformdirac{
\omega^{\alpha \beta}_D= \{Z^\alpha,Z^\beta \}^{\star_D}_M.} The
deformation consist in the fact that the symbol $F(Z^\alpha)$ is
actually a function on $\hbar$ such that $\lim_{\hbar \to 0}
F(Z^\alpha;\hbar) = f(Z^\alpha)$ and the Moyal-Dirac bracket
$\lim_{\hbar \to 0} \{F,G\}^{\star_D}_M = \{f,g\}_{D}.$

\noindent [Remark: In Refs.
\refs{\KrivoruchenkoTGuno,\KrivoruchenkoTGdos} the deformed Dirac
bracket $\{F,G\}^{\star_D}_M$ is usually written as the
anti-symmetric part of $\star_D$: $F(Z^\alpha) \wedge
G(Z^\alpha).$ A symmetric part also can be defined.]

The skew-projection of this Moyal product on the physical phase
space ${\cal Z}_P^{\cal R}$ is given by

\eqn\moyalreduc{
 \big(F \star_D  G \big) (Z^\alpha_S)=
F(Z^\alpha_{1S}) \exp\bigg\{{i\hbar\over 2}
\buildrel{\leftrightarrow}\over {\cal P}_D\bigg\} G(Z^\alpha_{2S})
\bigg|_{Z_{1S}^\alpha=Z_{2S}^\alpha=Z^\alpha_S},} where
\eqn\diracoperdos{ \buildrel{\leftrightarrow}\over {\cal P}_D =
\omega^{\alpha \beta}_D {\buildrel{\leftarrow}\over{\partial}
\over \partial Z^\alpha} \cdot
{\buildrel{\rightarrow}\over{\partial} \over
\partial Z^\beta}\bigg|_{Z=Z_S} =\buildrel{\leftrightarrow}\over {\cal P}.}
Morover, we know that the projection
 \eqn\deformdiracfdos{
\omega^{\alpha \beta}_D\bigg|_{{\cal Z}_P^{\cal R}}=
\{Z^\alpha_S,Z^\beta_S \}^{\star_D}_M= \{Z^\alpha_S,Z^\beta_S
\}_M.}

Now it is an easy matter to define the Wigner function. Assume
$\widehat{\rho}$ to be the density operator of the quantum state.
Then according to the general formula \sedos\ the function
$\rho(Z^\alpha)$ corresponding to $\widehat{\rho}$ reads
$$
{\rho}(Z^\alpha)= W^{-1}(\widehat{\rho})= {\rm Tr} \bigg(
\widehat{\Omega}(Z^\alpha) \widehat{\rho} \bigg).
$$
\eqn\secinco{ = \int d^{D} \xi\exp \bigg\{ -{i\over \hbar} \xi^\mu
\Pi_\mu \bigg\} \bigg\langle X^\mu + {{\xi}^\mu \over 2} \bigg|
\widehat{\rho} \bigg|X^\mu - {{\xi}^\mu \over 2} \bigg\rangle .}
Then the {\it Wigner function} ${\rho}^{W} (Z^\alpha)$ is defined
by a simple modification of Eq. \secinco. Namely,
$$
\rho^{W}(Z^\alpha) := \int d^{D} ({\xi \over 2 \pi \hbar}) \exp
\bigg\{ - { i \over \hbar} \xi^\mu \Pi_\mu \bigg\} \bigg\langle
{X}^\mu + {{\xi}^\mu \over 2} \bigg| \widehat{\rho} \bigg|{X}^\mu
- {{\xi}^\mu \over 2} \bigg\rangle.
$$
In particular for the pure state $\widehat{\rho} = |\Psi\rangle
\langle \Psi |$ we get
$$ \rho^{W}(Z^\alpha) = \int d^{D}({\xi \over 2 \pi \hbar})
\exp \bigg\{ - { i \over \hbar} {\xi}^\mu \Pi_\mu \bigg\}
\Psi^*\big(X^\mu - {{\xi}^\mu \over 2}\big) \Psi\big(X^\mu +
{{\xi}^\mu \over 2}\big), $$ where $\Psi({X}^\mu)=\langle X^\mu|
\Psi \rangle$ stands for $| \Psi \rangle$ in the Schr\"odinger
representation.

Given $\rho^{W}$ one can use Eq. \weylc\ to find the corresponding
density operator $\widehat{\rho}$

\eqn\otres{ \widehat{\rho} = \int d^{2D} Z \prod_{I=1}^{2m} \delta
[\Phi_I(Z^\alpha)] \sqrt{\det{\bf C}} \ \rho^{W}(Z^\alpha)
\widehat{\Omega}(Z^\alpha).} Consequently, the average value
$\langle \widehat{F} \rangle$ reads
$$
\langle \widehat{F} \rangle = { {\rm Tr} \big( \widehat{\rho}
\widehat{F} \big) \over {\rm Tr} \big( \widehat{\rho} \big).}
$$
Then
$$
\langle \widehat{F} \rangle  = { \int d^{2D} Z \ \sqrt{\det{\bf
C}} \prod_{I=1}^{2m} \delta [\Phi_I(Z^\alpha)]  \
\rho^{W}(Z^\alpha) \ {\rm Tr} \big( \widehat{\Omega}(Z^\alpha)
\widehat{F} \big) \over \int d^{2D} Z \ \sqrt{ \det{\bf C}}
\prod_{I=1}^{2m} \delta [\Phi_I(Z^\alpha)]
  \ \rho^{W}(Z^\alpha)}
$$
$$
 = { \int d^{2D} Z \ \sqrt{\det{\bf C}}
\prod_{I=1}^{2m} \delta [\Phi_I(Z^\alpha)]  \ \rho^{W}(Z^\alpha)
W^{-1} \big( \widehat{F} \big)(Z^\alpha) \over \int d^{2D}Z \
\sqrt{\det{\bf C}} \prod_{I=1}^{2m} \delta [\Phi_I(Z^\alpha)] \
\rho^{W}(Z^\alpha)}
$$
\eqn\ocuatro{ = { \int d^{2(D-m)} Z\ \sqrt{\det{\bf C}}
 \ \rho_S^{W}(Z^\alpha)
F_S(Z^\alpha) \over \int d^{2(D-m)}Z \ \sqrt{\det{\bf C}} \
\rho_S^{W}(Z^\alpha)}.}

\vskip 1truecm
\subsec{Correlation Functions}

Here we are going to present the simple example definition of the
correlation functions within the deformation quantization
formalism. Namely, we give the Green (Wightman) functions of any
pair of functions $\langle F(Z^\alpha) \star_D G(Z^\beta) \rangle$
as follows

\eqn\correlatone{ \langle F(Z^\alpha) \star_D  G(Z^\alpha) \rangle
= { \int d^{2D}Z \ \sqrt{\det{\bf C}} \ \prod_{I=1}^{2m} \delta
[\Phi_I(Z^\alpha)] \rho^{W}(Z^\alpha) F(Z^\alpha) \star_D
G(Z^\alpha) \over \int d^{2D}Z \ \sqrt{\det{\bf C}}
\prod_{I=1}^{2m} \delta [\Phi_I(Z^\alpha)] \ \rho^{W}(Z^\alpha)}.}
After integration over the variables involved in the constraints
we have \eqn\correlationtwo{\langle F(Z^\alpha) \star_D
G(Z^\alpha) \rangle = { \int d^{2(D-m)}Z \ \sqrt{\det{\bf C}} \
\rho^{{W}}_S(Z^\alpha) F_S(Z^\alpha) \star_D G_S(Z^\alpha) \over
\int d^{2(D-m)}Z \ \sqrt{\det{\bf C}} \ \rho_S^{W}(Z^\alpha)}.} In
the next section we will implement these results to quantize the
relativistic free particle in the light-cone gauge.

\vskip 2truecm
\newsec{Deformation Quantization of the Free Relativistic Particle in the Light-cone Gauge}

The purpose of this section we will apply the WWM-formalism of
second class constrained systems, surveyed in section 3, to the
case of the free relativistic particle. As it is well known in the
literature, the relativistic particle in the light-cone gauge (see
for instance, \refs{\barton}) can be quantized through the usual
Dirac prescription of quantization for the physical degrees of
freedom. We will show that this formalism can be consistently
applied to the relativistic particle. Before that we want to
describe the procedure to determine the dynamics on a second class
constrained system.

\vskip 1truecm
\subsec{Systems with Time-dependent Constraints}

Now we briefly overview the description of second class
constrained systems which does depend explicitly on time $\tau$.
This is precisely the case for all systems which are
$\tau$-reparametrization invariant like the case of the
relativistic particle. The canonical Hamiltonian analysis,
including the Hamiltonian evolution, is quite involved by this
fact and its implementation is quite non-trivial. To overcome
these difficulties a procedure based in geometrical analysis was
proposed in Ref. \tuckey\ and further applied to the relativistic
particle in an arbitrary electromagnetic background in Ref.
\refs{\evans}.

In general this gauge fixing problem defines a second class
constrained system with two possible time-dependent constraints
given by: \eqn\scceu{ \Phi_1=p^2 + m^2 =0, \ \ \ \ \ \ \
\Phi_2=X^+ - {1 \over m^2}p^+ \tau, \ \ \ \ \ ({\bf L})} for the
{\it light-cone} ${\bf L}$ gauge. The ${\bf L}$ gauge will be
crucial for the deformation quantization of the relativistic
particle.

This formalism allows to find a set of canonical variables for the
physical phase space ${\cal Z}_P^{\cal R}$, whose time evolution
can be described by an evolution equation of the type \eqn\eveqnu{
{dF \over dt} = {\partial F \over
\partial t} + \{F,H_D\}_D,} for any $F$ and where the original Hamiltonian
is modified in the form
\eqn\hamidirac{ H_D= H + K.} Here $K$ satisfies the following
differential equation \eqn\conditu{ Y= -\omega_{\alpha \beta}
{\partial Z^\alpha \over \partial \tau} {\partial Z^\beta \over
\partial \xi^a} d \xi^a = dK.}
In the previous equation $\omega_{\alpha \beta}$ is the symplectic
form in ${\cal Z}_P$ and $\xi^a$ are the local coordinates in the
physical phase space ${\cal Z}_P^{\cal R}$.

From this formalism it can be obtained the natural choice of the
physical coordinates, found in Sec. 2, is precisely: $\xi^a= (x^-,
p^+, \vec{x}_T, \vec{p}_T)$.  These physical coordinates satisfy
the following Dirac brackets:
$$
\{x^-, p_-\}_D=1, \ \ \ \ \ \ \ \{x^j,p_j\}_D=
\{X^j(\tau),\Pi_j(\tau)\}_D = \delta^i_j,
$$
where $p_- = -p^+,$ and $p_j = p^j$. If one solves the
differential equation \conditu, we find that $Y=dK= -{1\over
m^2}p^+ dp_+$ and therefore $K=\int Y={1\over m^2}p^+p^-$.
Moreover, the Hamiltonian $H_D$ is given \eqn\hamil{ H_D = {1\over
m^2}p^+p^- =\sum_{j=1}^{D-2}{p_j^2 + m^2 \over 2m^2}.} Obviously,
this expression coincides with the total Hamiltonian in Eq. \ham.

\vskip 1truecm
\subsec{WWM Formalism}

In the previous section we saw as the reduction to the physical
phase space keeping invariant the Hamiltonian dynamics is
justified by the formalism, at least locally in the phase space.
This is precisely what we need in order to apply the skew-gradient
projection formalism revised in Sec. 3 (see Eqs. \dinamica\ and
\dinamicados).

According to Sec. 2 and 3, the reduced phase space ${\cal
Z}_P^{\cal R}$ of the free relativistic particle in the light-cone
gauge is described by the (independent) variables
$(x^-,p^+,\vec{X}, \vec{\Pi}_T)$ or
$(x^-,p^+,\vec{x}_T,\vec{p}_T)$. Thus this reduced space ${\cal
Z}_P^{\cal R} = \IR^{2(D-1)}$ is endowed with the symplectic
two-form

\eqn\sdos{ \omega_{\cal R} = d p_- \wedge dx^- + \sum_{j=1}^{D-2}
dp_j \wedge dx^j.}

 This is precisely the pull-back of the
symplectic form $\omega$ of extended phase space, i.e.,
$\omega_{\cal R}= \phi^*(\omega)$. Here $\phi$ of the embedding of
the submanifold ${\cal Z}_P^{\cal R}$ into the unconstrained phase
space ${\cal Z}_P$.

 For the present case we have two constraints
\scceu. These constraints are of the second class since the
determinant of the ${\bf C}$-matrix is non-vanishing, i.e.
$\det(C_{IJ})\not=0$. Remember that locally it can be written as
$C_{IJ}=I_{IJ}$ and therefore $\det(C_{IJ})=1$.

Let $F = F(X^\mu,\Pi_\mu)$ be a function on the phase space ${\cal
Z}_P$. Then according to the Weyl rule \weylc\ we assign the
following operator  $\widehat{F}$ corresponding to $F$
$$
\widehat{F} = W(F)= \int  d^{D} X d^{D}( {\Pi \over 2 \pi \hbar})
\ \sqrt{\det {\bf C}}  \delta[\Phi_1(X^\mu,\Pi_\mu)]
\delta[\Phi_2(X^\mu,\Pi_\mu)] \ F(X^\mu,\Pi_\mu)
\widehat{\Omega}(X^\mu,\Pi_{\mu}),
$$
where the measures of the integrals are given by $d^{D}X = d X^0
\cdots dX^{D-1}$ and $d^{D}\Pi = d \Pi^0 \cdots d \Pi^{D-1}.$

Integration over the constraints reexpress the Weyl rule in ${\cal
Z}_P^{\cal R}$ in the form \eqn\ssiete{\widehat{F} = W(F)= \int
{dx^- dp^+ \over 2 \pi \hbar} d^{D-2} X_T d^{D-2}( {\Pi_T \over 2
\pi \hbar}) F(x^-,\vec{X}_T,p^+,\vec{\Pi}_T)
\widehat{\Omega}(x^-,\vec{X}_T,p^+,\vec{\Pi}_T),} where
$F(x^-,\vec{X}_T,p^+,\vec{\Pi}_T)=
F[(X^\mu,\Pi_\mu)_S]=F_{S}(X^\mu,\Pi_\mu)$ is the skew-gradient
projection of the symbol on the constrained phase space ${\cal
Z}_P$ and similarly for the SW quantizer $
\widehat{\Omega}(x^-,\vec{X}_T,p^+,\vec{\Pi}_T)=
\widehat{\Omega}[(X^\mu,\Pi_\mu)_S]=
\widehat{\Omega}_{S}(X^\mu,\Pi_\mu)$. Then Eq. \socho\ reduces to
$$
\widehat{\Omega}(x^-,\vec{X}_T,p^+,\vec{\Pi}_T) = \int d \xi^-
d^{D-2} \xi_T \exp \bigg\{ -{i\over \hbar} \big( - \xi^- p^+ +
\vec{\xi}_T \cdot \vec{\Pi}_T \big) \bigg\}
$$
$$
\times \bigg|x^- - {\xi^- \over 2}, \vec{X}_T - {\vec{\xi}_T \over
2} \bigg{\rangle} \bigg{\langle} \vec{X}_T + {\vec{\xi}_T \over
2}, x^- + {\xi^- \over 2}\bigg|
$$
\eqn\socho{ = \int d ({\eta^+\over 2 \pi \hbar}) d^{D-2}({\eta_T
\over 2 \pi \hbar}) \exp \bigg\{ - {i \over \hbar} \big( - x^-
\eta^+ + \vec{\eta}_T \cdot \vec{X}_T \big) \bigg\} \bigg| p^+ +
{\eta^+ \over 2}, \vec{\Pi}_T + {\vec{\eta}_T \over 2}
\bigg\rangle \bigg\langle \vec{\Pi}_T - {\vec{\eta}_T \over 2},
p^+ - {\eta^+ \over 2} \bigg|, } where $d^{D-2} \xi_T \equiv d
\xi^1 \dots d \xi^{D-2},$ $ d^{D-2}({\eta_T \over 2 \pi \hbar})
\equiv d ({\eta^1 \over 2 \pi \hbar}) \dots d ({\eta^{D-2} \over 2
\pi \hbar}),$ $\vec{\xi}_T \cdot \vec{\Pi}_T \equiv
\sum_{j=1}^{D-2} \xi^j \Pi^j$ and $\vec{\eta}_T \cdot \vec{X}_T
\equiv \sum_{j=1}^{D-2} \eta^j X^j.$ In the above expressions the
states in the reduced (physical) Hilbert space ${\cal H}_{phys}$
defined in the following way
$$
|x^-,\vec{X}_T \rangle := |x^- \rangle \otimes \bigg(
\bigotimes_{j=1}^{D-2} |X^j \rangle \bigg), \ \ \ \ \ \ \ \ |p^+,
\vec{\Pi}_T \rangle := |p^+ \rangle \otimes \bigg(
\bigotimes_{j=1}^{D-2} |\Pi^j \rangle \bigg).
$$
The SW quantizer has the same important properties \snueve\ on the
reduced phase space:

\eqn\snueve{ \big(
\widehat{\Omega}(x^-,\vec{X}_T,p^+,\vec{\Pi}_T)\big)^{\dag} =
\widehat{\Omega}(x^-,\vec{X}_T,p^+,\vec{\Pi}_T),} \eqn\setenta{
{\rm tr} \big(\widehat{\Omega}(x^-,\vec{X}_T,p^+,\vec{\Pi}_T)\big)
= 1,}
$$ {\rm tr} \big(\widehat{\Omega}(x^-,\vec{X}_T,p^+,\vec{\Pi}_T)
\widehat{\Omega}({x'}^-,{\vec{X}'}_T,{p'}^+,{\vec{\Pi}'_T})\big)
$$
\eqn\lastproperties{ = \delta (x^- -{x'}^-) \delta ({p^+ -{p'}^+
\over 2 \pi \hbar}) \delta[\vec{X}_T-\vec{X'}_T] \delta
[{\vec{\Pi}_T - \vec{\Pi'}_T \over 2 \pi \hbar}],} where ${\rm
tr}$ is the trace over ${\cal H}_{phys}$.

Multiplying Eq. \ssiete\ by
$\widehat{\Omega}(x^-,\vec{X}_T,p^+,\vec{\Pi}_T)$ and taking the
trace on the reduced Hilbert space, the reduced version of Eq.
\sedos\ is given by

\eqn\sedos{ F(x^-,\vec{X}_T,p^+,\vec{\Pi}_T) = {\rm tr}
\big[\widehat{\Omega}(x^-,\vec{X}_T,p^+,\vec{\Pi}_T) \widehat{F}
\big].}  Now, let $F = F(x^-,\vec{X}_T,p^+,\vec{\Pi}_T)$ and $G=
G(x^-,\vec{X}_T,p^+,\vec{\Pi}_T)$ be elements of $C^\infty({\cal
Z}_P^{\cal R})[[\hbar]]$ and let $\widehat{F}= W(F)$ and
$\widehat{G} =W(G)$ be their corresponding operators. The function
on ${\cal Z}_P^{\cal R}$ that corresponds to the product
$\widehat{F} \widehat{G}$ is denoted by $F \star G$ and takes the
form

\eqn\setres{ \big(F \star G \big) [x^-,\vec{X}_T,p^+,\vec{\Pi}_T]
:= W^{-1}(\widehat{F} \widehat{G})= {\rm Tr} \big[
\widehat{\Omega}(x^-,\vec{X}_T,p^+,\vec{\Pi}_T) \widehat{F}
\widehat{G} \big].}

Remember from Eq. \diracoperdos\ that
$\buildrel{\leftrightarrow}\over {\cal P}_D$ on ${\cal Z}_P^{\cal
R}$ reduces to $\buildrel{\leftrightarrow}\over {\cal P}$.
Substituting Eqs. \ssiete\ and \socho\ into \setres\ and
performing straightforward computations we can reexpress \setres\
as \eqn\moyalproyect{ \big(F \star G \big)
(x^-,\vec{X}_T,p^+,\vec{\Pi}_T) = F(x^-,\vec{X}_T,p^+,\vec{\Pi}_T)
\exp\bigg\{{i\hbar\over 2} \buildrel{\leftrightarrow}\over {\cal
P}\bigg\} G(x^-,\vec{X}_T,p^+,\vec{\Pi}_T),} where
$$
\buildrel{\leftrightarrow}\over {\cal P} =
\buildrel{\leftrightarrow}\over {\cal P}_\pm +
\buildrel{\leftrightarrow}\over {\cal P}_T
$$
\eqn\secuatro{ = \bigg({{\buildrel{\leftarrow}\over
{\partial}}\over
\partial p^{+}} {{\buildrel{\rightarrow}\over {\partial}}\over
\partial x^{-}} - {{\buildrel{\leftarrow}\over {\partial}}\over
\partial x^{-}} {{\buildrel{\rightarrow}\over {\partial}}\over
\partial p^{+}}\bigg) + \sum_{j=1}^{D-2}
\bigg({{\buildrel{\leftarrow}\over {\partial}}\over \partial
X^{j}} {{\buildrel{\rightarrow}\over {\partial}}\over \partial
\Pi^{j}} - {{\buildrel{\leftarrow}\over {\partial}}\over \partial
\Pi^{j}} {{\buildrel{\rightarrow}\over {\partial}}\over \partial
X^{j}}\bigg).}

Now we proceed to find the Wigner function.  Thus the density
operator $\widehat{\rho}$ of the quantum state of a the
relativistic particle is associated through Weyl correspondence
\sedos\ to the the function $\rho(x^-,\vec{X}_T,p^+,\vec{\Pi}_T)$
$$
{\rho}(x^-,\vec{X}_T,p^+,\vec{\Pi}_T)= W^{-1}(\widehat{\rho})=
{\rm tr} \bigg( \widehat{\Omega}(x^-,\vec{X}_T,p^+,\vec{\Pi}_T)
\widehat{\rho} \bigg)
$$
\eqn\secinco{ = \int d\xi^-  d^{D-2} \xi_T \exp \bigg\{ -{i\over
\hbar} \big( - \xi^- p^+ +  \vec{\xi}_T \cdot \vec{\Pi}_T \big)
\bigg\} \bigg\langle \vec{X}_T + {\vec{\xi}_T \over 2}, x^- +
{\xi^- \over 2} \bigg| \widehat{\rho} \bigg| x^- - {\xi^- \over
2}, \vec{X}_T - {\vec{\xi}_T \over 2} \bigg\rangle.} Then  as we
have seen the {\it Wigner function}
${\rho}^{W}(x^-,\vec{X}_T,p^+,\vec{\Pi}_T)$ is a slightly
modification of Eq. \secinco.  It is given by
$$
\rho^{W}(x^-,\vec{X}_T,p^+,\vec{\Pi}_T) := \int d ({\xi^- \over 2
\pi \hbar})d^{D-2} ({\xi \over 2 \pi \hbar}) \exp \bigg\{ - { i
\over \hbar} \big( - \xi^- p^+ +  \vec{\xi}_T  \cdot \vec{\Pi}_T
\big)\bigg\}
$$
\eqn\seseis{ \times \bigg\langle \vec{X}_T + {\vec{\xi}_T \over
2}, x^- + {\xi^- \over 2} \bigg| \widehat{\rho} \bigg| x^- -
{\xi^- \over 2}, \vec{X}_T - {\vec{\xi}_T \over 2} \bigg\rangle .}
For the pure state $\widehat{\rho} = |\Psi\rangle \langle \Psi |$
it can be written as
$$
\rho^{W}(x^-,\vec{X}_T,p^+,\vec{\Pi}_T) = \int d ({\xi^- \over 2
\pi \hbar})d^{D-2}({\xi_T \over 2 \pi \hbar})  \exp \bigg\{ - { i
\over \hbar} \big( - \xi^- p^+ +  \vec{\xi}_T \cdot \vec{\Pi}_T
\big)\bigg\}
$$
\eqn\sesiete{ \times \Psi^*\big(x^- - {\xi^- \over 2}, \vec{X}_T -
{\vec{\xi}_T \over 2}\big) \Psi\big(x^- + {\xi^- \over 2},
\vec{X}_T + {\vec{\xi}_T \over 2}\big),} where
$\Psi(x^-,\vec{X}_T)= \langle x^-,\vec{X}_T|\Psi\rangle.$

For completeness we will write down the equations in terms of
variables $(x^-,p^+, \vec{x}_T,\vec{p}_T$) one has
$$
\widehat{\Omega}(x^-,\vec{x}_T,p^+,\vec{p}_T) = \int d \xi^-
d^{D-2} \xi_T \exp \bigg\{ -{i\over \hbar} \bigg( - \xi^- p^+ +
\vec{\xi}_T \cdot \vec{p}_T \bigg) \bigg\}
$$
$$
\times \bigg|x^- - {\xi^- \over 2}, \vec{x}_T - {\vec{\xi}_T \over
2} \bigg\rangle \bigg\langle \vec{x}_T  + {\vec{\xi}_T \over 2},
x^- + {\xi^- \over 2} \bigg|
$$
\eqn\seocho{ = \int d ({\eta^+\over 2 \pi \hbar})  d ({\eta_T\over
2 \pi \hbar}) \ \exp \bigg\{ - {i \over \hbar} \bigg( - x^- \eta^+
+ \vec{\eta}_T \cdot \vec{x}_T \bigg) \bigg\} \bigg| p^+ + {\eta^+
\over 2}, \vec{p}_T + {\vec{\eta}_T \over 2} \bigg\rangle
\bigg\langle \vec{p}_T - {\vec{\eta}_T \over 2}, p^+ - {\eta^+
\over 2} \bigg|,} where $\vec{\xi}_T \cdot \vec{p}_T \equiv
\sum_{j=1}^{D-2} \xi^j p^j,$ $\vec{\eta}_T \cdot \vec{x}_T \equiv$
$\sum_{j=1}^{D-2}$ $ \eta^j x^j.$

Then the Moyal $*$-product in terms of these variables reads as
\eqn\otromoyal{ \big(F_1 \star  F_2\big)
(x^-,\vec{x}_T,p^+,\vec{p}_T) = F_1(x^-,\vec{x}_T,p^+,\vec{p}_T)
\exp\bigg\{{i\hbar\over 2} \buildrel{\leftrightarrow}\over {\cal
P} \bigg\} F_2(x^-,\vec{x}_T,p^+,\vec{p}_T),} with \eqn\senueve{
\buildrel{\leftrightarrow}\over {\cal P} :=
\bigg({{\buildrel{\leftarrow}\over {\partial}}\over
\partial p^{+}} {{\buildrel{\rightarrow}\over {\partial}}\over
\partial x^{-}} - {{\buildrel{\leftarrow}\over {\partial}}\over
\partial x^{-}} {{\buildrel{\rightarrow}\over {\partial}}\over
\partial p^{+}}\bigg) + \sum_{j=1}^{D-2}
\bigg({{\buildrel{\leftarrow}\over {\partial}}\over
\partial x^{j}} {{\buildrel{\rightarrow}\over {\partial}}\over
\partial p^{j}} - {{\buildrel{\leftarrow}\over {\partial}}\over
\partial p^{j}} {{\buildrel{\rightarrow}\over {\partial}}\over
\partial x^{j}}\bigg).}

Finally, for the Wigner function one obtains
$$
\rho^{W}(x^-,\vec{x}_T,p^+,\vec{p}_T) = \int d ({\xi^- \over 2 \pi
\hbar}) d^{D-2} ({\xi_T \over 2 \pi \hbar}) \exp \bigg\{ - { i
\over \hbar} \bigg( - \xi^- p^+ +  \vec{\xi}_T \cdot \vec{p}_T
\bigg)\bigg\}
$$
\eqn\ouno{ \times \bigg\langle \vec{x}_T + {\vec{\xi}_T \over 2},
x^- + {\xi^- \over 2} \bigg| \widehat{\rho} \bigg| x^- - {\xi^-
\over 2}, \vec{x}_T - {\vec{\xi}_T \over 2} \bigg\rangle } and in
the case of the pure state we get
$$
\rho^{W}(x^-,\vec{x}_T,p^+,\vec{p}_T) = \int d ({\xi^- \over 2 \pi
\hbar}) d^{D-2} ({\xi_T \over 2 \pi \hbar})  \exp \bigg\{ - { i
\over \hbar} \bigg( - \xi^- p^+ +  \vec{\xi}_T \cdot \vec{p}_T
\bigg)\bigg\}
$$

\eqn\odos{  \times \Psi^*\big(x^- - {\xi^- \over 2}, \vec{x}_T -
{\vec{\xi}_T \over 2}\big) \Psi\big(x^- + {\xi^- \over 2},
\vec{x}_T + {\vec{\xi}_T \over 2}\big).}

The application of the inverse Weyl correspondence \ssiete\ allows
to find the  density operator $\widehat{\rho}$ starting from a
given Wigner function $\rho^{W}$. This is given by

\eqn\otres{ \widehat{\rho} = \int d x^- d ({p^+\over 2 \pi \hbar})
d^{D-2} x_T d^{D-2}( {p_T \over 2 \pi \hbar}) \
\rho^{W}(x^-,\vec{x}_T,p^+,\vec{p}_T)
\widehat{\Omega}(x^-,\vec{x}_T,p^+,\vec{p}_T).} Thus, the
correlation function of the operator $\widehat{F}$ is given by
$$
\langle \widehat{F} \rangle = { {\rm tr} \big( \widehat{\rho}
\widehat{F} \big) \over {\rm tr} \big( \widehat{\rho} \big)}
$$
$$
= { \int d x^- d ({p^+ \over 2 \pi \hbar}) d^{D-2}x_T d^{D-2}({p_T
\over 2 \pi \hbar}) \rho^{W}(x^-,\vec{x}_T,p^+,\vec{p}_T) \ {\rm
Tr} \big( \widehat{\Omega}(x^-,\vec{x}_T,p^+,\vec{p}_T)
\widehat{F} \big) \over \int d x^- d({p^+ \over 2 \pi \hbar})
d^{D-2}x_T d^{D-2} ({p_T \over 2 \pi \hbar})
\rho^{W}(x^-,\vec{x}_T,p^+,\vec{p}_T)}
$$
\eqn\ocuatro{ = { \int d x^- d ({p^+ \over 2 \pi \hbar})
d^{D-2}x_T d^{D-2}({p_T \over 2 \pi \hbar}) \
\rho^{W}(x^-,\vec{x}_T,p^+,\vec{p}_T) W^{-1} \big( \widehat{F}
\big)(x^-,\vec{p}_T) \over \int d x^- d ({p^+ \over 2 \pi \hbar})
d^{D-2}x_T d^{D-2}({p_T \over 2 \pi \hbar})
\rho^{W}(x^-,\vec{x}_T,p^+,\vec{p}_T)}.}

\vskip 1truecm
\subsec{Wigner Function of the Ground State}

The Wigner function $\rho^{W}_0$ of the free particle is defined
by

\eqn\oocho{ p^j \star \rho^W_0=0 \ \ \ \  {\rm and} \ \ \ \ p^+
\star \rho^{W}_0 = 0, } for $j=1, \dots , D-2$.

Then after expanding the star product \otromoyal\ we have
\eqn\onueve{ p^j \rho^{W}_0 = 0, \ \ \ \ \ \ p^+ \rho^{W}_0 = 0, }
for $j=1, \dots , D-2$. The general real solution of Eq. \onueve\
satisfying also Eqs. $(I.6)$ and $(I.7)$ (see appendix I) reads

\eqn\noventa{ \rho^W_0 =C \delta (p^1) \dots \delta(p^{D-2})
\delta(p^+),} where $C$ is a positive integration constant i.e.,
$C>0$.

Observe that $\rho^{W}_0$ is defined by Eqs. (I.6), (I.7) and
\oocho\ uniquely up to an arbitrary real positive constant factor
$C > 0$. This fact can be interpreted in deformation quantization
formalism as the uniqueness of the ground state.

\vskip 1truecm
\subsec{Green Functions}

In the present subsection we are going to compute the correlation
functions in the context of the deformation quantization
formalism. We would like to find the Green (Wightman) functions
with $F(Z^\alpha) = X^j(\tau)$ and $G(Z^\alpha)= X^k(\tau')$. Then
we will compute from \correlationtwo\ $\langle X^j(\tau) \star
X^k(\tau ') \rangle$. Substituting it into the definition
\ocuatro\ and after integration over the constraints we find

\eqn\cienvtres{ \langle X^j(\tau) \star X^k(\tau ') \rangle = {
\int d x^- d ({p^+ \over 2 \pi \hbar}) d^{D-2}x_T d^{D-2}({p_T
\over 2 \pi \hbar}) \rho^{W}_0(x^-,\vec{x}_T,p^+,\vec{p}_T)
X^j(\tau) \star X^k(\tau ') \over \int d x^- d ({p^+ \over 2 \pi
\hbar}) d^{D-2}x_T d^{D-2}({p_T \over 2 \pi \hbar})
\rho^{W}_0(x^-,\vec{x}_T,p^+,\vec{p}_T)}.} where $\rho^{W}_0$ is
given by Eq. \noventa\ and $X^j(\tau)$ is given by Eq. \doce.
After some computations we have
$$
\langle x^j \star p^k \rangle = \delta_{jk} {i \hbar \over 2} = -
\langle p^j \star x^k \rangle
$$
\eqn\cienvseis{ \langle p^j \star p^k \rangle = 0,  \ \ \ \ \ \ \
\langle x^j \star x^k \rangle = \delta_{jk} \langle x^j x^k
\rangle.} After a bit of algebra one finally find that the two
point correlation functions are given by \eqn\cienvsietelast{
\langle X^j(\tau) \star X^k(\tau ') \rangle = \langle x^j x^k
\rangle + {i \hbar \delta_{jk}\over 2 m^2} (\tau '- \tau).} In the
next section we describe the deformation quantization of our
particle in a general electromagnetic background. This description
will be in the context of the procedure of sections 3 and 4. We
will see that the quantization can be carried over in a natural
way into this context.

\vskip 2truecm
\newsec{Deformation Quantization of a Relativistic Particle in
a General Electromagnetic Background}

In the present section we will apply the WWM-formalism for second
class constraints discussed in section 3, to the case of the
relativistic particle in an arbitrary electromagnetic background.
We use also the result to deal time-dependent constraints from
symplectic geometrical analysis \refs{\evans,\tuckey}. Thus the
particle in an electromagnetic background can be properly
canonically quantized in a time-dependent gauge like the temporal
gauge or the light-cone gauge. The involved electromagnetic field
is non-dynamical and only will works as a general background, thus
the number of degrees of freedom do not change and the physical
phase space is still our friend ${\cal Z}_P^{\cal R}$. The
corresponding action in the conformal gauge is given by

\eqn\lagrangcoupledone{ S = \int_{L} d \tau \bigg[{m^2\over 2}
\eta_{\mu \nu}{dX^{\mu} \over d \tau} {dX^{\nu}\over d\tau} - {1
\over 2}\bigg] - e \int_L  d \tau \ A_{\mu}(X^\rho) {dX^{\mu}
\over d \tau}.} From this, the canonical momentum is easily
computed and it yields

\eqn\lagrangcoupledtwo{ P_\mu = m^2 \dot{X}_\mu - e A_\mu=p_\mu
-eA_\mu.}

The Poisson bracket of these dynamical variables is given by
\eqn\newbracket{ \{X^\mu,P^\nu\}_P = \eta^{\mu \nu}.}

The presence of an electromagnetic field, as a background, does
not modifies the fact that we are dealing with a second class
constrained system whose constraints are time-dependent. This
leads again to a time-dependent gauge-fixing procedure.
Fortunately, as we see in the previous section, there is a
formulation explored in Ref. \refs{\tuckey,\evans} for
time-dependent constraints.

In general terms, this gauge fixing problem defines a second class
constrained system with two time-dependent constraints given by:
\eqn\scce{ \Phi_1=(P+eA)^2 + m^2 =0, \ \ \ \ \ \ \ \ \Phi_2=X^+ -
{1 \over m^2}P^+ \tau, \ \ \ \ \ ({\bf L}).}

In Ref. \evans, it was shown than even introducing the
electromagnetic field as an arbitrary background, there is a
Hamiltonian evolution  (see, Eq. \conditu) description on the
system on the physical phase space ${\cal Z}_P^{\cal R}$. Also in
this case the formalism allows to find a set of canonical
variables for the physical phase space ${\cal Z}_P^{\cal R}$ are
precisely: $(X^-, P^+, \vec{X}_T, \vec{P}_T)$, whose time
evolution can be described by an equation of the type \eveqnu.
Thus, essentially the number of degrees of freedom are the same
than for the free particle. These variables also satisfy the
following Dirac brackets

\eqn\newrelaciones{ \{X^-, P_-\}_D=1, \ \ \ \ \ \ \ \{X^j,P_j\}_D=
\{X^j(\tau),\Pi_j(\tau)\}_D = \delta^i_j.}

Moreover, from Eq. \conditu\ we can compute $Y=dK= -{1\over
m^2}p^+ dp_+$ and therefore $K=\int Y={1\over m^2}p^+p^-$, with
$p^\pm=P^\pm+eA^+,$ we have that the total Hamiltonian is

$$
H_D= K = \int Y ={1 \over m^2}(p^++e A^+)(p^-+e A^-)
$$
\eqn\laye{={1 \over 2m^2} \sum_{j=1}^{D-2} \big[ (P^j+ eA^j)^2 +
m^2 \big],} which is the natural generalization of \ham.

 \vskip
1truecm

\subsec{Deformation Quantization of the Relativistic Particle in
an Arbitrary Electromagnetic Background}

In this subsection we are going to describe the deformation
quantization for a relativistic particle in an arbitrary
electromagnetic background. This is a straightforward
generalization of the discussion of subsection 4.2.  We have seen
that the number of degrees of freedom is exactly the same as the
free particle. The reason of that is the fact that electromagnetic
field as a background does not introduce new degrees of freedom.
Consequently, the Hilbert space description is also the same. Thus
the quantization follows straightforward by substituting Poisson
brackets by the Dirac one and take the Dirac prescription of
quantization. Then we only describe the necessary modifications to
quantize by deformation.

Let us consider $F = F[X^\mu,P_\mu]$ be a function on the phase
space ${\cal Z}_P$. Then according to the Weyl rule \weylc\ we
assign the following operator  $\widehat{F}$ corresponding to $F$
\eqn\weylcorrespwithA{ \widehat{F} = W(F)= \int  d^{D} X d^{D}( {P
\over 2 \pi \hbar}) \delta[(P+eA)^2+m^2] \delta[X^+ -{1 \over
m^2}P^+ \tau] \sqrt{\det {\bf C}} \ F[X^\mu,P_\mu]
\widehat{\Omega}[X^\mu,P_{\mu}].}

Integration over the irrelevant dof's  have essentially the same
description as for the free particle described in section 4. Thus
we have

\eqn\ssiete{ \widehat{F} = W(F)= \int {dX^- dP^+ \over 2 \pi
\hbar} d^{D-2} X_T d^{D-2}( {P_T \over 2 \pi \hbar})
F[X^-,\vec{X}_T,P^+,\vec{P}_T]
\widehat{\Omega}[X^-,\vec{X}_T,P^+,\vec{P}_T],} where

$F[X^-,\vec{X}_T,P^+,\vec{P}_T]$ and
$\widehat{\Omega}[X^-,\vec{X}_T,P^+,\vec{P}_T]$ are the
skew-gradient projections on ${\cal Z}_P^{\cal R}$. The projected
SW quantizer is given by
$$
\widehat{\Omega}[X^-,\vec{X}_T,P^+,\vec{P}_T] = \int d \xi^-
d^{D-2} \xi_T \exp \bigg\{ -{i\over \hbar} \big( - \xi^- P^+ +
\vec{\xi}_T \cdot \vec{P}_T \big) \bigg\}
$$
$$
\times \bigg|X^- - {\xi^- \over 2}, \vec{X}_T - {\vec{\xi}_T \over
2} \bigg{\rangle} \bigg{\langle} \vec{X}_T + {\vec{\xi}_T \over
2}, x^- + {\xi^- \over 2}\bigg|
$$
\eqn\socho{ = \int d ({\eta^+\over 2 \pi \hbar}) d^{D-2}({\eta_T
\over 2 \pi \hbar}) \exp \bigg\{ - {i \over \hbar} \big( - X^-
\eta^+ + \vec{\eta}_T \cdot \vec{X}_T \big) \bigg\} \bigg| P^+ +
{\eta^+ \over 2}, \vec{P}_T + {\vec{\eta}_T \over 2} \bigg\rangle
\bigg\langle \vec{P}_T - {\vec{\eta}_T \over 2}, P^+ - {\eta^+
\over 2} \bigg| } with the obvious notation $ \vec{\xi}_T \cdot
\vec{P}_T \equiv \sum_{j=1}^{D-2} \xi^j P^j$ and $\vec{\eta}_T
\cdot \vec{X}_T \equiv \sum_{j=1}^{D-2} \eta^j X^j.$

Thus, besides the complications on the gauge fixing procedure,
which is encoded in the delta functions from \weylcorrespwithA,
otherwise the procedure is identical as the free particle in the
light-cone gauge described in section 4 and it will be not
repeated here. In the next section we couple the particle with a
dynamical electromagnetic field, which will be a more interesting
system to study.

\vskip 2truecm

\newsec{Deformation Quantization of the Charged Point Particle in a Dynamical
 Electromagnetic Field}

In this section we will consider the deformation quantization for
the relativistic point particle interacting with a dynamical
electromagnetic field. For simplicity we shall restricted our
selves to the case $D=4$, but, the analysis can be easily extended
to higher dimensions. The corresponding action in the conformal
gauge is given by

\eqn\lagrangcoupledonedos{ S = \int_{L} d \tau \bigg[{m^2\over 2}
\eta_{\mu \nu}{dX^{\mu} \over d \tau} {dX^{\nu}\over d\tau} - {1
\over 2}\bigg] - e  \int_L  d \tau \ A_{\mu}(X^\rho) {dX^{\mu}
\over d \tau}+ \int d^4x \bigg\{- {1\over 4e^2}F^{\mu \nu} F_{\mu
\nu} - {1 \over 2} \zeta({\cal S}[ A^\mu]) \bigg\},} where $\zeta$
is a Lagrange multiplier and the last term in the action is the
gauge fixing term. We will consider the Lorentz gauge
$G=\partial_\rho A^\rho =0$ and the light-cone gauge $G=A^+=0$.

The momenta are easily computed and them yields

\eqn\lagrangcoupledtwodos{ P_\mu = {\partial L \over \partial
\dot{X}^\mu}= m^2 \dot{X}_\mu - e A_\mu= p_\mu-eA_\mu, \ \ \ \ \ \
\pi^\mu= {\partial L \over \partial (\partial_tA)}=- F^{\mu 0} -
\zeta \eta^{\mu 0} G[A^\mu].} with $\mu, \nu=0,1,2,3$ and $\pi^i =
-E^i$ being the components of the electric field.

Deformation quantization of classical electromagnetic field in the
Coulomb gauge was discussed in Ref. \campos. In this section we
will adopt rather a covariant description, thus we will employ the
general formalism from section 3.

The Poisson brackets for the electromagnetic field is given by
$$
\{A_\mu(\vec{x},t), \pi_{\nu}(\vec{y},t) \}_P = \eta_{\mu \nu}
\delta(\vec{x} - \vec{y}),
$$
\eqn\seocho{ \{A_\mu(\vec{x},t), A_\nu(\vec{y},t) \}_P = 0 =
\{\pi_\mu(\vec{x},t),\pi_\nu(\vec{y},t)\}_P.}

The plane-wave expansion of the field variables reads

\eqn\aes{ A^\mu(\vec{x},t) = { 1\over (2 \pi)^{3/2}} \int d^3k
\bigg({ \hbar \over 2 \omega(k)}\bigg)^{1/2}  \bigg(
a^{\mu}(\vec{k},t) \exp\big(i \vec{k} \cdot \vec{x} \big) +
a^{*\mu}(\vec{k},t) \exp \big(-i \vec{k} \cdot \vec{x} \big)
\bigg),}

\eqn\senuevedos{ \pi^\mu(\vec{x},t)  = { 1\over (2 \pi)^{3/2}}
\int d^3k \ i \bigg({\hbar \omega(\vec{k}) \over 2}\bigg)^{1/2}
\bigg( a^{\mu}(\vec{k},t)\exp\big(i \vec{k} \cdot \vec{x} \big) -
a^{*\mu}(\vec{k},t) \exp \big(-i \vec{k} \cdot \vec{x} \big)
\bigg),} where $\omega(\vec{k})=|\vec{k}|$, $a^\mu(\vec{k},t)=
\sum_{\lambda=0}^3 a(\vec{k},\lambda,t)
\varepsilon^\mu(\vec{k},\lambda)$ and
$a(\vec{k},\lambda,t)=a(\vec{k},\lambda) \exp[-i \omega(k)t]$.
Notice that the vector potential \aes\ can be decomposed into
positive and negative frequency modes: $A^\mu(\vec{x},t)=
A^{+\mu}(\vec{x},t) + A^{-\mu}(\vec{x},t).$

Here $\varepsilon^\mu(\vec{k},\lambda)$ are four polarization
vectors which satisfy the following relations \eqn\relaciones{
\varepsilon^\mu(\vec{k},\lambda)
\varepsilon_\mu(\vec{k},\lambda')= \eta_{\lambda \lambda'}, \ \ \
\ \sum_{\lambda=0}^3 \eta_{\lambda \lambda}
\varepsilon^\mu(\vec{k},\lambda) \varepsilon^\nu(\vec{k},\lambda)=
\eta^{\mu \nu},} and \eqn\otra{k \cdot \varepsilon(\vec{k},1)= k
\cdot \varepsilon(\vec{k},2) =0, \ \ \ \ k \cdot
\varepsilon(\vec{k},0) =- k \cdot \varepsilon(\vec{k},3).}

The mode functions $a^\mu(\vec{k},\lambda)$ can be obtained from
the vector potential and its conjugated variable
$$
a^{\mu} (\vec{k},t) = {1 \over (2 \pi)^{3/2} ( 2 \hbar
\omega(\vec{k}))^{1/2}} \int d^3x \exp \big( -i \vec{k} \cdot
\vec{x} \big) \bigg( \omega(\vec{k}) A^\mu(\vec{x},t) - i \pi^\mu
(\vec{x},t) \bigg),
$$
\eqn\odoss{ a^{*\mu} (\vec{k},t) = {1 \over (2 \pi)^{3/2} ( 2
\hbar \omega(\vec{k}))^{1/2}} \int d^3x \exp \big(i \vec{k} \cdot
\vec{x} \big) \bigg( \omega(\vec{k}) A^\mu(\vec{x},t) + i \pi^\mu
(\vec{x},t) \bigg). } They satisfy the Poisson brackets
$$
\{a_{\mu}(\vec{k},t), a^*_{\nu}(\vec{k}',t) \}_P = - {i \over \hbar} \eta_{\mu \nu}
\delta(\vec{k} - \vec{k}'), $$

\eqn\otres{ \{a_{\mu}(\vec{k},t), a_{\nu}(\vec{k}',t) \}_P = 0 =
\{a^*_{\mu}(\vec{k},t),a^*_{\nu}(\vec{k}',t)\}_P.}

Then in Feynman gauge ($\zeta=1$) and in the Lorentz gauge the
Hamiltonian of the particle and the electromagnetic field reads
$$
H_T = H_C + \sum_{I=1}^{2} \lambda_I \Phi_I(X^\mu,P_\mu) - {1
\over 2} \int d^3x \bigg( \pi^\mu \pi_\mu - {1 \over 2}
\partial_k A_\mu \partial^k A^\mu \bigg) + e \int_L d \tau
A_\mu(X^\rho) {d X^\mu \over d \tau}
$$
$$ =  H_C + \sum_{I=1}^2 \lambda_I \Phi_I(X^\mu,P_\mu)+ \int d^3k
\sum_{\lambda=1}^2 \hbar \omega(\vec{k})
a^*(\vec{k},\lambda)a(\vec{k},\lambda)
$$
\eqn\manyhami{ + \int d^3k \hbar
\omega(\vec{k})\big[a^*(\vec{k},3) a(\vec{k},3) - a^*(\vec{k},0)
a(\vec{k},0) \big].}

Notice that this Hamiltonian is not positive-definite and
therefore there will be not have a ground state. In order to make
this Hamiltonian positive one usually impose the Lorentz gauge
condition on the positive frequency modes of the gauge field

\eqn\lorentzgaugeclasss{
\partial_\mu {A}^{+\mu}(\vec{x},t)=0.} Equivalently we
have \eqn\samething{ \sum_{\lambda=0}^3 k \cdot
\varepsilon(\vec{k},\lambda) a(\vec{k},\lambda)=0.} If we use the
transversality conditions for the massless photon \otra, we get
\eqn\longitudinal{ \big[{a}(\vec{k},3) - {a}(\vec{k},0)\big] =0.}
It is easy to see that this condition removes the ambiguity in the
Hamiltonian \manyhami.

\vskip 1truecm
\subsec{ The Stratonovich-Weyl Quantizer for the Complete System}

Consider now the Weyl quantization of the complete system.  To
this end we deal with the relevant fields at the time $t=0$. Let
$F=F[X^\mu,P_\mu,A^\mu,\pi_\mu]$ be an element of $C^\infty( {\cal
Z}_P \times {\cal Z}_M)[[\hbar]]$. The Weyl rule assign to the
functional $F$ the following operator $\widehat{F}$
$$
 \widehat{F} = W(F[X^\mu,P_\mu,A^\mu,\pi_\mu]):= \int
d^{4} X d^{4}( {P \over 2 \pi \hbar})  {\cal D} A {\cal D} ({\pi
\over 2 \pi \hbar}) \ \sqrt{\det {\bf C}}
$$
\eqn\onuevedos{
 \times  \delta[\Phi_1(X^\mu,P_\mu)]
\delta[\Phi_2(X^\mu,P_\mu)] \ \delta[G(A^\mu)]
 \ F[X^\mu,P_\mu,A^\mu,\pi_\mu] \widehat{\Omega}[X^\mu,P_{\mu},A^\mu,\pi_\mu], }
where $G[A^\mu]=0$ represents the Lorentz or light-cone gauge, and
$\widehat{\Omega}[X^\mu,P_{\mu},A^\mu,\pi_\mu]$ is the
Stratonovich-Weyl quantizer for the whole system.

\vskip 1truecm \subsec{Lorentz Gauge}

First we consider as field variables of the electromagnetic field,
the oscillator variables $a$ and $a^*$. The Weyl rule reads
$$
 \widehat{F} = W(F[X^\mu,P_\mu,a^\mu,a^{*\mu}]):= \int
d^{4} X d^{4}( {P \over 2 \pi \hbar})  {\cal D}a{\cal D} a^* \
\sqrt{\det {\bf C}}
$$
\eqn\onuevedostres{
 \times  \delta[\Phi_1(X^\mu,P_\mu)]
\delta[\Phi_2(X^\mu,P_\mu)] \ \delta[G(a^\mu,a^{*\mu})]
 \ F[X^\mu,P_\mu,a^\mu,a^{*\mu}] \widehat{\Omega}[X^\mu,P_{\mu},a^\mu,a^{*\mu}]. }

In particular in the light-cone gauge for the particle sector and
the Lorentz gauge \longitudinal, for the electromagnetic sector we
have
$$
 \widehat{F} = W(F[X^\mu,P_\mu,a^\mu,a^{*\mu}]):= \int
d^{D} X d^{D}( {P \over 2 \pi \hbar})  {\cal D}a{\cal D} a^* \
\sqrt{\det {\bf C}}
$$
\eqn\onuevedostres{
 \times  \delta[(P+eA)^2 +m^2]
\delta[X^+ - {1\over m^2}P^+ \tau] \ \delta[a(\vec{k},3) -
a(\vec{k},0)] F[X^\mu,P_\mu,a^\mu,a^{*\mu}] \widehat{\Omega}
 [X^\mu,P_{\mu},a^\mu,a^{*\mu}]. }
As we have seen the procedure to gauge fixing the part of particle
is easy to implement. The part corresponding to the
electromagnetic field is more involved and we will concentrate on
it. 

The Stratonovich-Weyl quantizer is given by
$$
\widehat{\Omega}[X^\mu,P_\mu,A^\mu,\pi_\mu] = \int d^{4} \xi {\cal
D}\lambda(\vec{x}) \exp \bigg\{ -{i\over \hbar} {\xi}^\mu P_\mu
-{i \over \hbar}\int d^{3}x \lambda^\mu(\vec{x}) \pi_\mu(\vec{x})
\bigg\}
$$
$$
\times \bigg|X^\mu - {\xi^\mu \over 2}, A^\mu -{\lambda^\mu \over
2} \bigg{\rangle} \bigg{\langle} A^\mu + {\lambda^\mu \over
2},X^\mu + {{\xi}^\mu \over 2}\bigg|
$$
$$
 = \int d^4({\eta\over 2 \pi \hbar}) {\cal D}
\lambda^{*\mu}(\vec{x}) \exp \bigg\{ - {i \over \hbar} \eta^\mu
X_\mu -{i \over \hbar}\int d^{3}x \lambda^{*\mu}(\vec{x})
A_\mu(\vec{x}) \bigg\}
$$
\eqn\sochoo{ \times  \bigg| P^\mu + {\eta^\mu \over 2}, \pi^\mu
-{\lambda^{*\mu} \over 2} \bigg{\rangle} \bigg{\langle} \pi^\mu
+{\lambda^{\mu*}\over 2}, P^\mu - {\eta^\mu \over 2} \bigg|.}
The structure of the extended Fock space ${\cal H}_p \otimes {\cal
F}_M$ is given by
$$ |X^\mu, A^\mu \rangle = |X^\mu \rangle \otimes |A^\mu
\rangle
$$
where $|X^\mu \rangle = |X^+,X^-,\vec{X}_T \rangle = |X^+\rangle
\otimes |X^- \rangle \otimes |\vec{X}_T \rangle$ and $|A^\mu
\rangle =  |a(\vec{k},0), a(\vec{k},3), \vec{a}_T \rangle =
|a(\vec{k},0) \rangle \otimes |a(\vec{k},3) \rangle \otimes
|\vec{a}_T \rangle$.

The commutation relations are given by
$$
[\widehat{a}^\mu (\vec{k},t), \widehat{a}^{*\nu} (\vec{k}',t)]=
\eta^{\mu \nu} \delta(\vec{k} - \vec{k}'),
$$
\eqn\nrelaciones{[\widehat{a}^\mu (\vec{k},t), \widehat{a}^{\nu}
(\vec{k}',t)]=0=[\widehat{a}^{*\mu} (\vec{k},t),
\widehat{a}^{*\nu} (\vec{k}',t)].}

Following a similar analysis to the description of the
supersymmetric Weyl correspondence from Ref. \imelda, we get
\eqn\pproducto{ \widehat{\Omega}[X^\mu,P_\mu,a,a^*]=
\widehat{\Omega}[X^\mu,P_\mu] \otimes \widehat{\Omega}[a,a^*].}
These operators, of course, can be skew-gradient projected as
follows

\eqn\proyeccion{ \widehat{\Omega}_S[X^\mu,P_\mu,a,a^*]=
\widehat{\Omega}_S[X^\mu,P_\mu] \otimes
\widehat{\Omega}_S[a,a^*].}

\vskip .5truecm
\noindent{\it The Star-Product}

 The Moyal $\star$-product in the complete system can be
constructed in a similar way as for the free case. Let
$F[X^\mu,P_\mu,a,a^*]$ and $G[X^\mu,P_\mu,a,a^*]$ be two
functionals on ${\cal Z}_P\times {\cal Z}_M$ and let $\widehat{F}$
and $\widehat{G}$ be their corresponding Weyl operators, then

\eqn\moyalcasi{
 \big(F \star  G \big)[X^\mu,P_\mu,a,a^*] = F[X^\mu,P_\mu,a,a^*]
  \exp\bigg({i\hbar\over 2}
\buildrel{\leftrightarrow} \over {\cal P}_{pM}\bigg)
G[X^\mu,P_\mu,a,a^*],} where \eqn\suma{ \buildrel{\leftrightarrow}
\over {\cal P}_{pM}= \buildrel{\leftrightarrow} \over {\cal P}_p +
\buildrel{\leftrightarrow} \over {\cal P}_M.} Here
\eqn\poissonpart{ \buildrel{\leftrightarrow}\over {\cal P}_p :=
 \sum_{\mu=0}^{3}
\bigg({{\buildrel{\leftarrow}\over {\partial}}\over \partial
X^{\mu}} {{\buildrel{\rightarrow}\over {\partial}}\over
\partial P_{\mu}} - {{\buildrel{\leftarrow}\over {\partial}}\over
\partial P^{\mu}} {{\buildrel{\rightarrow}\over {\partial}}\over
\partial X_{\mu}}\bigg)} and

$$
\buildrel{\leftrightarrow}\over {\cal P}_M := \sum_{\lambda=0}^{3}
\int d^{3}k \bigg({{\buildrel{\leftarrow}\over {\delta}}\over
\delta a(\vec{k},\lambda)} {{\buildrel{\rightarrow}\over
{\delta}}\over \delta a^{*}(\vec{k},\lambda)} -
{{\buildrel{\leftarrow}\over {\delta}}\over \delta \
a^{*}(\vec{k},\lambda)} {{\buildrel{\rightarrow}\over
{\delta}}\over \delta a^(\vec{k},\lambda)}\bigg)
$$
$$
=\sum_{\lambda=1}^{2} \int d^{3}k
\bigg({{\buildrel{\leftarrow}\over {\delta}}\over \delta
a(\vec{k},\lambda)} {{\buildrel{\rightarrow}\over {\delta}}\over
\delta a^{*}(\vec{k},\lambda)} - {{\buildrel{\leftarrow}\over
{\delta}}\over \delta \ a^{*}(\vec{k},\lambda)}
{{\buildrel{\rightarrow}\over {\delta}}\over \delta
a^(\vec{k},\lambda)}\bigg)
$$
$$
+ \int d^{3}k \bigg({{\buildrel{\leftarrow}\over {\delta}}\over
\delta a(\vec{k},0)} {{\buildrel{\rightarrow}\over {\delta}}\over
\delta a^{*}(\vec{k},0)} - {{\buildrel{\leftarrow}\over
{\delta}}\over \delta \ a^{*}(\vec{k},0)}
{{\buildrel{\rightarrow}\over {\delta}}\over \delta
a^(\vec{k},0)}\bigg) $$ \eqn\oppoisson{ + \int d^{3}k
\bigg({{\buildrel{\leftarrow}\over {\delta}}\over \delta
a(\vec{k},3)} {{\buildrel{\rightarrow}\over {\delta}}\over \delta
a^{*}(\vec{k},3)} - {{\buildrel{\leftarrow}\over {\delta}}\over
\delta \ a^{*}(\vec{k},3)} {{\buildrel{\rightarrow}\over
{\delta}}\over \delta a^(\vec{k},3)}\bigg).}

\vskip .5truecm
\noindent{\it The Wigner Functional for the Complete System}

Now we are going to implement the gauge constraints at the quantum
level. These are the particle and the gauge field parts. The
particle part is the usual light-cone gauge, while the
electromagnetic part will be the quantum version of the Lorentz
gauge \eqn\lorentzgauge{
\partial_\mu \widehat{A}^{+\mu}(\vec{x},t)| \Phi \rangle=0.} Equivalently we
have \eqn\samething{ \sum_{\lambda=0}^3 k \cdot
\varepsilon(\vec{k},\lambda) \widehat{a}(\vec{k},\lambda) |\Phi
\rangle=0.} States $|\Phi \rangle$ are factorized as $|\Phi
\rangle = |\Phi_P \rangle \otimes |\Phi_U \rangle$, where $|\Phi_P
\rangle$ are the physical states and $|\Phi_U \rangle$ are
unphysical ones. Eq. \samething\ leads to \eqn\longitudinal{
\big[\widehat{a}(\vec{k},3) - \widehat{a}(\vec{k},0)\big]|\Phi
\rangle = \big[\widehat{a}(\vec{k},3) -
\widehat{a}(\vec{k},0)\big]|\Phi_U \rangle =0.} Here we have used
the well known transversality conditions for the massless photon
\otra.

Equivalently one can use the light-cone gauge for the
electromagnetic part \eqn\lcgauge{ \widehat{A}^+(\vec{x},t)|\Phi
\rangle=0.} In the procedure of quantization there will be used
both of them.

Now we are going to compute the physical Winger functional of the
ground state. For the composed system it was shown in Ref.
\imelda, that the Wigner function can be factorized as

\eqn\factorizacion{\rho^W_0(X^\mu,P_\mu,a,a^*) =
\rho^W_0(X^\mu,P_\mu) \cdot  \rho^W_0(a,a^*).} The skew-gradient
projected is the physical Wigner functional
\eqn\factorizaciond{\rho^W_{S0}(X^\mu,P_\mu,a,a^*) =
\rho^W_{S0}(X^\mu,P_\mu) \cdot  \rho^W_{S0}(a,a^*).}

For the particle case will have $\rho^W_{S0}(X^\mu,P_\mu)=
\rho^W_{0}(X^-,\vec{X}_T, P^+,\vec{P}_T)$ as we got previously
(see, Eq. \noventa). Thus the Wigner functional
$\rho^W_{S0}(X^\mu,P_\mu) \cdot \rho^W_{0}(a,a^*)$ will be a
solution of the following systems of equations
$$
P^j \star \rho^W_0 =0, \ \ \ \ \ \ \ \ \  P^+ \star \rho^W_0=0,
$$
\eqn\constraints{ \vec{a}_T(\vec{k},j) \star \rho^W_0=0, \ \ \
 \ \ \ \ \big[a(\vec{k},3) - a(\vec{k},0) \big] \star \rho^W_0 =0.}
Taking $\rho^W_0(a,a^*) =\rho^W_0(P) \cdot \rho^W_0(U)$ and using
the Moyal product \poissonpart\ and \oppoisson\ we get

$$
P^j \rho^W_0 =0, \ \ \ \ \ \ \ \ \ P^+ \rho^W_0=0,
$$
\eqn\const{ {a}^j_T(\vec{k})  \rho^W_0(P) +{1 \over 2} {\delta
\rho^W_0(P) \over \delta a^{*j}_T(\vec{k})} =0, \ \ \ \ \ \
\big[a(\vec{k},3) - a(\vec{k},0) \big] \rho^W_0(U)=0.} The
solution is as follows to all these constraints is given by the
product of $\rho^W_{0}(X^-,\vec{X}_T, P^+,\vec{P}_T)=
\delta(\vec{P}_T) \delta(P^+)$ and $\rho^W_{0}(a,a^*) =\rho^W_0(P)
\cdot \rho^W_0(U)= \exp \bigg \{ - 2 \sum_{j=1}^{2}\int d^{3}k
a^{*j}_T(\vec{k}) a^j_T(\vec{k}) \bigg\} \delta \big[a(\vec{k},3)
- a(\vec{k},0) \big].$ Then we have

\eqn\wignerfunction{ \rho^{W}_{0}(X^-,\vec{X}_T,
P^+,\vec{P}_T,a,a^*) =C \exp \bigg \{ - 2 \sum_{j=1}^{2}\int
d^{3}k a^{*j}_T(\vec{k}) a^j_T(\vec{k}) \bigg\} \delta
\big[a(\vec{k},3) - a(\vec{k},0) \big] \delta (\vec{P}_T)
\delta(P^+),} where $C
> 0$. This Wigner function composed by the Wigner function of
the particle plus that of an infinite set of oscillators resembles
very much that of the bosonic string \strings.

Thus, once we have the Wigner functional of the ground state, the
correlation functions of gauge invariant operators $\widehat{\cal
O}$ can be computed by
$$
\langle \widehat{\cal O} \rangle = { {\rm Tr} \big(
\widehat{\rho}_{phys} \widehat{\cal O} \big) \over {\rm Tr} \big(
\widehat{\rho}_{phys} \big)}
$$
$$
= { \int d X^- d P^+ d^{2}X_T d^{2}P_T {\cal D}a_T {\cal D} a^*_T
\rho^{W}_0[X^-,\vec{X}_T,P^+,\vec{P}_T,\vec{a}_T,\vec{a}^*_T] \
{\rm Tr} \big(
\widehat{\Omega}[X^-,\vec{X}_T,P^+,\vec{P}_T,\vec{a}_T,\vec{a}^*_T]
\widehat{\cal O} \big) \over \int d X^- d P^+ d^{2}X_T d^{2} P_T
\rho^{W}_0[X^-,\vec{X}_T,P^+,\vec{P}_T,\vec{a}_T,\vec{a}^*_T]}
$$
\eqn\ocuatro{ = { \int d X^- d P^+ d^{2}X_T d^{2}P_T  {\cal D}a_T
{\cal D} a^*_T\
\rho^{W}_0[X^-,\vec{X}_T,P^+,\vec{P}_T,\vec{a}_T,\vec{a}^*_T]
{\cal O}_T[X^-,\vec{X}_T,P^+,\vec{P}_T,\vec{a}_T,\vec{a}^*_T]
\over \int d X^- d P^+ d^{2}X_T d^{2}P_T
\rho^{W}_0[X^-,\vec{X}_T,P^+,\vec{P}_T,\vec{a}_T,\vec{a}^*_T]}.}

\vskip .5truecm
\noindent{\it Gauge Invariant Reduction}

After integrating out the spurious degrees of freedom with the
uses of \oppoisson\ we have
$$
\widehat{F} = W(F)= \int {dX^- dP^+ \over 2 \pi \hbar} d^{2} X_T
d^{2}( {P_T \over 2 \pi \hbar}) {\cal D} a_T {\cal D}a^*_T \
\sqrt{\det {\bf C}}
$$
\eqn\ssietedos{
 \times F[X^-,\vec{X}_T,P^+,\vec{P}_T,\vec{a}_T,\vec{a}^*_T]
\widehat{\Omega}[X^-,\vec{X}_T,P^+,\vec{P}_T,\vec{a}_T,\vec{a}^*_T]}
where $F[X^-,\vec{X}_T,P^+,\vec{P}_T,\vec{a}_T,\vec{a}^*_T]=
F[[X^\mu,P_\mu,a^\mu,a^*_\mu]_S]=
F_{S}[X^\mu,P_\mu,a^\mu,a^*_\mu]$ is the skew-gradient projected
symbol on the reduced phase space ${\cal Z}^{\cal R}_P \times
{\cal Z}^{\cal R}_M$ and similarly for the Stratonovich-Weyl
quantizer
$\widehat{\Omega}[X^-,\vec{X}_T,P^+,\vec{P}_T,\vec{a}_T,\vec{a}^*_T]=
\widehat{\Omega}[[X^\mu,P_\mu,a^\mu,a^*_\mu]_S]=\widehat{\Omega}_{S}[X^\mu,P_\mu,a^\mu,a^*_\mu]$
which is given by
$$
\widehat{\Omega}[X^-,\vec{X}_T,P^+,\vec{P}_T,\vec{a}_T,\vec{a}^*_T]
= \widehat{\Omega}[X^-,\vec{X}_T,P^+,\vec{P}_T] \otimes
\widehat{\Omega}[\vec{a}_T,\vec{a}^*_T]
$$
$$
=\int d \xi^- d^{2} \xi_T  {\cal D}\lambda_T \exp \bigg\{ -{i\over
\hbar} \big( - \xi^- P^+ + \vec{\xi}_T \cdot \vec{P}_T \big) -{i
\over \hbar}\int d^{3}x \big(\vec{\lambda}_T \cdot \vec{a}^*_T
\big) \bigg\}
$$
$$
\times \bigg|X^- - {\xi^- \over 2}, \vec{X}_T - {\vec{\xi}_T \over
2}, \vec{a}_T -{\vec{\lambda}_T \over 2} \bigg{\rangle}
\bigg{\langle} \vec{a}_T +{\vec{\lambda}_T\over 2},\vec{X}_T +
{\vec{\xi}_T \over 2}, X^- + {\xi^- \over 2}\bigg|
$$
$$
= \int d ({\eta^+\over 2 \pi \hbar}) d^{2}({\eta_T \over 2 \pi
\hbar}) {\cal D} \lambda^*_T \exp \bigg\{ - {i \over \hbar} \big(
- X^- \eta^+ + \vec{\eta}_T \cdot \vec{X}_T \big) -{i \over
\hbar}\int d^{3}x \big(\vec{\lambda}^*_T \cdot \vec{a}_T
\big)\bigg\}
$$
\eqn\socho{
 \times \bigg| P^+ + {\eta^+ \over 2}, \vec{P}_T
+ {\vec{\eta}_T \over 2}, \vec{a}^*_T -{\vec{\lambda}^*_T \over 2}
\bigg{\rangle} \bigg{\langle} \vec{a}^*_T +{\vec{\lambda}^*_T\over
2}, \vec{P}_T - {\vec{\eta}_T \over 2}, P^+ - {\eta^+ \over 2}
\bigg| } with the obvious notation $\vec{\xi}_T \cdot \vec{P}_T
\equiv \sum_{j=1}^{2} \xi^j P^j$, $\vec{\eta}_T \cdot \vec{X}_T
\equiv \sum_{j=1}^{2} \eta^j X^j,$ $\vec{\lambda}_T \cdot
\vec{a}^*_T \equiv \sum_{j=1}^{2} \lambda^j a^{*j}$ and
$\vec{\lambda}^*_T \cdot \vec{a}_T \equiv \sum_{j=1}^{2}
\lambda^{*j} a^j.$

\vskip 1truecm \subsec{Light-cone Gauge}

Now we move from the field variables to the light-cone variables.
In particular in the light-cone gauge for both the particle and
the Maxwell gauge field. The gauge field $A^\mu$ can be decomposed
in $(A^+,A^-,\vec{A}_T)$. It will be more convenient to work in
the moment space representation $(A^+(p),A^-(p),\vec{A}_T)$. We
will work in the light-cone gauge $A^+(p)=0$. In this gauge fields
$A^-(p)$ are determined in terms of $p^+$ and the transverse
components $\vec{A}_T$ i.e., $A^-(p) ={1 \over P^+} \vec{p}_T
\cdot \vec{A}_T$. Thus for $p^2=0$ the gauge field is completely
determined by the two transverse degrees of freedom $\vec{A}_T$.
The light-cone gauge can be incorporated into the Weyl
correspondence as follows
$$
\widehat{F} = W(F[X^\mu,P_\mu,A^\mu,\pi_{\mu}]): = \int d^{D} X
d^{D}( {P \over 2 \pi \hbar})  {\cal D}A{\cal D} ( {\pi \over 2
\pi \hbar})  \ \sqrt{\det {\bf C}}
$$
\eqn\onuevedostres{ \times \delta[(P+eA)^2 +m^2] \delta[X^+ -
{1\over m^2}P^+ \tau] \ \delta[A^+] F[X^\mu,P_\mu,A^\mu,\pi_{\mu}]
 \widehat{\Omega}[X^\mu,P_{\mu},A^\mu,\pi_{\mu}]. }

\vskip .5truecm
\noindent{\it The Star-Product}

We start with the functionals $F[X^\mu,P_\mu,A^\mu,\pi_\mu]$ and
$G[X^\mu,P_\mu,A^\mu,\pi_\mu]$ defined on ${\cal Z}_P\times {\cal
Z}_M$ and let $\widehat{F}$ and $\widehat{G}$ be their
corresponding operators, then

$$
\big(F \star  G \big)[X^\mu,P_\mu,A^\mu,\pi_\mu]
$$
\eqn\otravezmoyal{
  = F[X^\mu,P_\mu,A^\mu,\pi_\mu]
  \exp\bigg({i\hbar\over 2}
\buildrel{\leftrightarrow} \over {\cal P}_{pM}\bigg)
G[X^\mu,P_\mu,A^\mu,\pi_\mu],}

\eqn\sumamas{ \buildrel{\leftrightarrow}\over {\cal P}_{pM} =
\buildrel{\leftrightarrow}\over {\cal P}_{p} +
\buildrel{\leftrightarrow}\over {\cal P}_{M},} where

\eqn\particuladost{
 \buildrel{\leftrightarrow}\over {\cal P}_{p}:=
 \sum_{\mu=0}^{3}
\bigg({{\buildrel{\leftarrow}\over {\partial}}\over \partial
X^{\mu}} {{\buildrel{\rightarrow}\over {\partial}}\over \partial
P_{\mu}} - {{\buildrel{\leftarrow}\over {\partial}}\over \partial
P^{\mu}} {{\buildrel{\rightarrow}\over {\partial}}\over \partial
X_{\mu}}\bigg),} and
$$
 \buildrel{\leftrightarrow}\over {\cal P}_{M}:=
\sum_{\mu=0}^{3} \int d^{3}x \bigg({{\buildrel{\leftarrow}\over
{\delta}}\over \delta A^{\mu}(\vec{x})}
{{\buildrel{\rightarrow}\over {\delta}}\over \delta
\pi_{\mu}(\vec{x})} - {{\buildrel{\leftarrow}\over {\delta}}\over
\delta \ \pi^{\mu}(\vec{x})} {{\buildrel{\rightarrow}\over
{\delta}}\over \delta A_\mu(\vec{x})}\bigg)
$$
\eqn\electcampo{ =\int d^{3}x \bigg({{\buildrel{\leftarrow}\over
{\delta}}\over \delta A^{+}(\vec{x})}
{{\buildrel{\rightarrow}\over {\delta}}\over \delta
\pi_{-}(\vec{x})} - {{\buildrel{\leftarrow}\over {\delta}}\over
\delta \ \pi^{+}(\vec{x})} {{\buildrel{\rightarrow}\over
{\delta}}\over \delta A_-(\vec{x})}\bigg) + \sum_{j=1}^2\int
d^{3}x \bigg({{\buildrel{\leftarrow}\over {\delta}}\over \delta
A^{j}(\vec{x})} {{\buildrel{\rightarrow}\over {\delta}}\over
\delta \pi_{j}(\vec{x})} - {{\buildrel{\leftarrow}\over
{\delta}}\over \delta \ \pi^{j}(\vec{x})}
{{\buildrel{\rightarrow}\over {\delta}}\over \delta
A_j(\vec{x})}\bigg).}

\vskip .5truecm
\noindent{\it The Wigner Functional for the Complete System}

Now we are going to compute the Winger functional of the ground
state. The Wigner functional we want to compute is
\eqn\factorizaciondos{\rho^W_0(X^\mu,P_\mu,a,a^*) =
\rho^W_{S0}(X^\mu,P_\mu) \cdot  \rho^W_0(a,a^*).} This is defined
by the conditions

$$
P^j \star \rho^W_0 =0, \ \ \ \ \ \ \ \ \  P^+ \star \rho^W_0=0,
$$
\eqn\masrestric{\vec{a}_T(\vec{p},j) \star \rho^W_0=0, \ \ \ \ \ \
\  A^+(\vec{p}) \star \rho^W_0 =0.} As before, take
$\rho^W_0(a,a^*)= \rho^W_0(P) \otimes \rho^W_0(U)$ and with aid of
the Moyal product
\particuladost\ and \electcampo\ we get
$$
P^j \rho^W_0 =0, \ \ \ \ \ \ \ \ \  P^+ \rho^W_0=0,
$$
\eqn\casilast{ {a}^j_T(\vec{p}) \rho^W_0(P) + {1\over 2} {\delta
\rho^W_0(P) \over \delta a^{*j}_T(\vec{p})} =0, \ \ \ \ \ \ \ \
A^+(\vec{p}) \rho^W_0(U)=0.}

The solution is as follows to all these constraints is given by

\eqn\wignerlc{\rho^{W}_0(X^-,\vec{X}_T, P^+,\vec{P}_T,a,a^*)  =C
\exp \bigg \{ - 2 \sum_{j=1}^{2} \int d^{3}p a^{*j}_T(\vec{p})
a^j_T(\vec{p}) \bigg\} \delta \big[A^+(\vec{p}) \big]
 \delta (\vec{P}_T)\delta(P^+),} with $C > 0$. The correlation
functions also can be computed in this gauge with this Wigner
function.

\vskip .5truecm
\noindent{\it Gauge Invariant Reduction}

After integrating out the spurious degrees of freedom it yields
$$
\widehat{F} = W(F)= \int {dX^- dP^+ \over 2 \pi \hbar} d^{2} X_T
d^{2}( {P_T \over 2 \pi \hbar}) {\cal D} A_T {\cal D}\pi_T \
\sqrt{\det {\bf C}}
$$
\eqn\ssietedos{
 \times F[X^-,\vec{X}_T,P^+,\vec{P}_T,\vec{A}_T,\vec{\pi}_T]
\widehat{\Omega}[X^-,\vec{X}_T,P^+,\vec{P}_T,\vec{A}_T,\vec{\pi}_T]}.
The Stratonovich-Weyl quantizer is given by
$$
\widehat{\Omega}[X^-,\vec{X}_T,P^+,\vec{P}_T,\vec{A}_T,\vec{\pi}_T]
= \widehat{\Omega}[X^-,\vec{X}_T,P^+,\vec{P}_T] \otimes
\widehat{\Omega}[\vec{A}_T,\vec{\pi}_T]
$$
$$
= \int d \xi^- d^{2} \xi_T  {\cal D}\lambda_T \exp \bigg\{
-{i\over \hbar} \big( - \xi^- P^+ + \vec{\xi}_T \cdot \vec{P}_T
\big) -{i \over \hbar}\int d^{3}x \vec{\lambda}_T(\vec{x}) \cdot
\vec{\pi}_T(\vec{x}) \bigg\}
$$
$$
\times \bigg|X^- - {\xi^- \over 2}, \vec{X}_T - {\vec{\xi}_T \over
2}, \vec{A}_T -{\vec{\lambda}_T \over 2} \bigg{\rangle}
\bigg{\langle} \vec{A}_T +{\vec{\lambda}_T\over 2},\vec{X}_T +
{\vec{\xi}_T \over 2}, X^- + {\xi^- \over 2}\bigg|
$$
$$
= \int d ({\eta^+\over 2 \pi \hbar}) d^{2}({\eta_T \over 2 \pi
\hbar}) {\cal D} \lambda^*_T \exp \bigg\{ - {i \over \hbar} \big(
- X^- \eta^+ + \vec{\eta}_T \cdot \vec{X}_T \big) -{i \over
\hbar}\int d^{3}x \vec{\lambda}^*_T(\vec{x}) \cdot
\vec{A}_T(\vec{x})\bigg\}
$$
\eqn\sochoo{
 \times \bigg| P^+ + {\eta^+ \over 2}, \vec{P}_T
+ {\vec{\eta}_T \over 2}, \vec{\pi}_T -{\vec{\lambda}^*_T \over 2}
\bigg{\rangle} \bigg{\langle} \vec{\pi}_T +{\vec{\lambda}^*_T\over
2}, \vec{P}_T - {\vec{\eta}_T \over 2}, P^+ - {\eta^+ \over 2}
\bigg|.}

 \vskip 2truecm
\newsec{Final Remarks}

In the present paper we have applied the WWM-formalism to quantize
the relativistic free particle and the relativistic particle in a
general electromagnetic background. We have used recent results
concerning the deformation quantization of second class
constrained systems
\refs{\LouisMartinezKW,\KrivoruchenkoTGuno,\KrivoruchenkoTGdos}.
We have shown that this formalism serves to quantize both kind of
systems in a way that resembles the Faddeev and Popov quantization
of a gauge theory through Feynman path integrals. It allows to
describe the deformation quantization of constrained systems in a
more geometric way. This approach also is useful to quantize by
deformation the charged particle interacting with a dynamical
electromagnetic field in the Lorentz and the light-cone gauges.

Moreover, we obtain that the WWM-formalism used, totally justifies
the deformation quantization of the relativistic particle in the
light-cone gauge. Which can be regarded also as the low energy
limit when the size of the string vanishes, i.e., $\alpha ' =
\ell_S^2 \to 0.$ Thus, our results are consistent with those of
\strings\ in this limit.

We have shown that deformation quantization of the relativistic
particle gives the same results as the canonical quantization and
path integral methods. Thus, this equivalence constitutes an
further evidence of the validity of these proposals
\refs{\LouisMartinezKW,\KrivoruchenkoTGuno,\KrivoruchenkoTGdos}
for systems with second class constraints. The Stratonovich-Weyl
quantizer, Weyl correspondence, Moyal product and the Wigner
function are obtained for all the analyzed systems.

The extension of the formalism to the superparticle described by
the supersymmetric action
$$
S_{SP} = \int_{L} d \tau \ \eta_{\mu \nu} \ \bigg({dX^{\mu} \over
d \tau} -i \overline{\theta}\Gamma^\mu{d \theta \over d
\tau}\bigg) \bigg({dX^{\nu} \over d \tau} -i
\overline{\theta}\Gamma^\nu{d \theta \over d \tau}\bigg).
$$
and more general systems like superstring theory is one of the
open problems that will be pursued in the near future. It is
interesting also to apply all these matters to more complicated
second class constrained systems as the BRST quantization in gauge
theories and Batalin-Vilkovisky quantization. It would be
interesting also to describe the deformation quantization of the
closed and open strings coupled with Neveu-Schwartz and
Ramond-Ramond fields. We hope to address some of these topics in
the future.

\vskip 2truecm

\centerline{\bf Acknowledgements}

We wish to thank Merced Montesinos for useful discussions. I.G.
and H.G.-C. are indebted Prof. M. Przanowski for much
encouragement. This paper was partially supported by a CONACyT
(M\'exico) grant 45713-F. Laura S\'anchez wish to thank Cinvestav,
Unidad Monterrey for its hospitality during her visit where part
of this work was done. This work represents a thesis for a
Bachelor degree submitted at the Facultad de Ciencias, UAE by
Laura S\'anchez. The research of I.G. was supported by a CONACyT
graduate fellowship. I.G. wants to thank Cinvestav Unidad
Monterrey for the generous hospitality. Finally, H.G.-C. wish to
express his enormous gratitude to Sra. Rosa Mar\'{\i}a
Compe\'an$^{\dag}$ for invaluable encouragement.

\vfill
\break
\appendix{I}{Wigner Functions and Pure States}

For the pure state the density operator is $\widehat{\rho} = |
\Psi \rangle \langle \Psi |.$ One can substitute it into Eq.
\otres, and with the aid of Eq. \seocho\ after some lengthy
algebra we get \eqn\ocinco{ | \Psi({y}^-,\vec{y}_T) |^2 = \int d
({p^+ \over 2 \pi \hbar}) d^{D-2}({p_T \over 2 \pi \hbar}) \
\rho^{W}({y}^-,\vec{y}_T,p^+,\vec{p}_T).} Here we assumed that
$\Psi({y}^-,\vec{y}_T) \not= 0.$ From this one can extract the
wave function $\Psi(x^-,\vec{x}_T)$ in terms of the corresponding
Wigner function $\rho^{W}$
$$
\Psi(x^-,\vec{x}_T) = {\cal N}^{-1} \exp \big\{ i \varphi \big\}
\int d ({p^+\over 2 \pi \hbar}) d^{D-2}({p_T \over 2 \pi \hbar}) \
\rho^{W}\big({x^-+ {y}^-\over 2},{\vec{x}_T+ \vec{y}_T\over
2},p^+,\vec{p}_T\big)
$$
\eqn\oseis{
 \times \exp \bigg\{ - {i \over \hbar} \big( -(x^- - {y}^-) p^+ + (\vec{x}_T - \vec{y}_T)
  \cdot
 \vec{p}_T \big) \bigg\} ,} where
$\exp(i \varphi)$ is a phase factor with $\varphi$ being a real
constant and

\eqn\normaliz{ {\cal N}= \bigg(\int d ({p^+ \over 2 \pi \hbar})
d^{D-2}({p_T \over 2 \pi \hbar})
\rho^{W}({y}^-,\vec{y}_T,p^+,\vec{p}_T) \bigg)^{1/2}.}

In terms of variables $\vec{X}_T(\tau)$ and $\vec{\Pi}_T(\tau)$
one has
$$
\Psi(x^-,\vec{X}_T) =  {\cal N}'^{-1} \exp \big\{ i \varphi \big\}
\int d ({p^+ \over 2 \pi \hbar}) d^{D-2}({\Pi_T \over 2 \pi
\hbar}) \ \rho^{W}\big({x^- + {y}^-\over 2},{\vec{X}_T +
\vec{Y}_T\over 2},p^+,\vec{\Pi}_T\big)
$$
\eqn\osiete{ \times \exp \bigg\{ - {i \over \hbar} \big( -(x^- -
{y}^-) p^+ + (\vec{X}_T(\tau) - \vec{Y}_T(\tau))\cdot \vec{\Pi}_T
\big) \bigg\},} where $\vec{X}_T(\tau)\cdot \vec{\Pi}_T(\tau)
\equiv \sum_{j=1}^{D-2} X^j \Pi^j$ and \eqn\normalizdos{ {\cal
N}'=\bigg(\int d ({p^+ \over 2 \pi \hbar}) d^{D-2} ({\Pi_T \over 2
\pi \hbar}) \rho^{W}\big({y}^-,\vec{X}_T, p^+,\vec{\Pi}_T)\big)
\bigg)^{1/2}.}

The natural question that arises is: at what extent the real
function $\rho^{W}(x^-,\vec{x}_T,p^+,\vec{p}_T)$ represents some
quantum state, {\it i.e.} it can be considered to be a Wigner
function. The necessary and sufficient conditions are

\eqn\lacan{ \int dx^- d({p^+ \over 2 \pi \hbar}) d^{D-2}x_T
d^{D-2}({p_T \over 2 \pi \hbar})
\rho^{W}(x^-,\vec{x}_T,p^+,\vec{p}_T) \big[ f^*
* f \big] (x^-,\vec{x}_T,p^+,\vec{p}_T) \geq 0,} for any $f \in C^{\infty}({\cal
Z}_P^{\cal R})[[\hbar ]]$ and

\eqn\lacanu{ \int dx^- d({p^+ \over 2 \pi \hbar}) d^{D-2}x_T
d^{D-2}({p_T \over 2 \pi \hbar})
\rho^{W}(x^-,\vec{x}_T,p^+,\vec{p}_T) > 0.}

In this appendix we study the situation in which a real function
$\rho^{W}(x^-,\vec{x}_T,p^+,\vec{p}_T),$ satisfying the positivity
conditions of Eqs. \lacan\ and \lacanu, represents the Wigner
function of a pure state.

In order to describe this situation in the case of a system of
particles we will use the results of Ref. \tata. In our case the
solution is quite similar. To begin with we denote
$$
\gamma(x^-,\vec{x}_T,{y}^-, \vec{y}_T):= \int d ({p^+ \over 2 \pi
\hbar}) d^{D-2} ({p_T \over 2 \pi \hbar}) \rho^{W}\big({x^- +
{y}^- \over 2}, {\vec{x}_T + \vec{y}_T \over 2},
p^+,\vec{p}_T\big)
$$

\eqn\ntres{ \times  \exp \bigg\{ {i \over \hbar}\big[- (x^- -
{y}^-)p^+ + (\vec{x}_T - \vec{y}_T) \cdot \vec{p}_T \big]
\bigg\}.} From Eq. \oseis\ it follows that if $\rho^{W}$ is the
Wigner function of the pure state $|\Psi \rangle \langle \Psi |$
then the functions $\gamma$ must satisfy the following equations
$$
{\partial^2 \ln \gamma(x^-,\vec{x}_T,{y}^-,\vec{y}_T) \over
\partial x^- \partial {y}^-} = {\partial^2 \ln
\gamma(x^-,\vec{x}_T,{y}^-,\vec{y}_T) \over \partial x^-
\partial
{y}^j}
$$
\eqn\ncuatro{ = {\partial^2 \ln
\gamma(x^-,\vec{x}_T,{y}^-,\vec{y}_T) \over \partial {x}^j
\partial {y}^-}= {\partial^2 \ln
\gamma(x^-,\vec{x}_T,{y}^-,\vec{y}_T) \over \partial {x}^j
\partial {y}^k}= 0}
for every $j,k= 1, \dots , D-2$.

Conversely, let $\gamma$ satisfies Eq. \ncuatro. The general
solution of \ncuatro\ can be factorized as follows \eqn\ncinco{
\gamma(x^-,\vec{x}_T,{y}^-,\vec{y}_T) = \Psi_1(x^-,\vec{x}_T)
\Psi_2({y}^-, \vec{y}_T).} As the function $\rho_{_W}$ is assumed
to be real we get from Eq. \ntres\

\eqn\nseis{ \gamma^*(x^-,\vec{x}_T,{y}^-,\vec{y}_T) =
\gamma({y}^-,\vec{y}_T, {x}^-,\vec{x}_T).} Consequently, Eq.
\ncinco\ has the form

\eqn\nsiete{ \gamma(x^-,\vec{x}_T,{y}^-,\vec{y}_T) = A
\Psi_1(x^-,\vec{x}_T) \Psi_1^*({y}^-,\vec{y}_T),} where, by the
assumption \lacanu, $A$ is a positive real constant. Finally,
defining $\Psi:= \sqrt{A} \Psi_1(x^-,\vec{x}_T)$ one obtains

\eqn\nsiete{ \gamma(x^-,\vec{x}_T,{y}^-,\vec{y}_T) =
\Psi(x^-,\vec{x}_T) \Psi^*({y}^-, \vec{y}_T).} Substituting $x^-
\mapsto x^- + {\xi^- \over 2},$ $\vec{x}_T \mapsto \vec{x}_T +
{\vec{\xi}_T\over 2},$ ${y}^- \mapsto x^- - {\xi^- \over 2},$
$\vec{x}_T \mapsto \vec{x}_T - {\vec{\xi}_T \over 2},$ multiplying
both sides by $\exp \big\{ - {i \over \hbar} (- \xi^- p^+ +
\vec{\xi}_T \cdot \vec{p}_T)\big\}$ and integrating with respect
to $ d({\xi^- \over 2 \pi \hbar}) d({ \xi \over 2 \pi \hbar})$ we
get exactly the relation \odos. This means that our function
$\rho_{_W}$ is the Wigner function of the pure state
$\Psi(x^-,\vec{x}_T)$.

In terms of variables $(x^-,\vec{X}_T,p^+, \vec{\Pi}_T)$ the
conditions \ncuatro\ read
$$
{\partial^2 \ln \gamma(x^-,\vec{X}_T,{y}^-,\vec{Y}_T) \over
\partial x^- \partial \tilde{x}^-} = {\partial^2 \ln
\gamma(x^-,\vec{X}_T,{y}^-,\vec{Y}_T)\over \partial x^- \partial
\vec{Y}_T }
$$
\eqn\nocho{ = {\partial^2  \ln
\gamma(x^-,\vec{X}_T,{y}^-,\vec{Y}_T)\over \partial \tilde{x}^-
\partial \vec{X}_T}= {\partial^2 \ln
\gamma(x^-,\vec{X}_T,{y}^-,\vec{Y}_T) \over \partial \vec{X}_T
\partial \vec{Y}_T}= 0.}

\listrefs

\end